%% file: main.tex
\documentclass[a4paper,11pt]{article}
\pdfoutput=1 

\usepackage{jheppub} 
\usepackage[]{physics}
\usepackage{amsmath}
\usepackage{breqn}
\usepackage{fancyvrb}
\usepackage{amstext} 
\usepackage{array}   
\newcolumntype{L}{>{$}l<{$}} 

\usepackage{braket}
\usepackage[T1]{fontenc} 
\usepackage{verbatim}
\usepackage{empheq} 
\usepackage[toc,page]{appendix}
\usepackage[utf8]{inputenc}
\usepackage{gensymb}
\usepackage{physics}
 \usepackage{url}
 \usepackage{amssymb}
    \usepackage{amsthm}
    \usepackage{amsfonts}
    \usepackage{subcaption} 
\usepackage{braket}
\usepackage{fancyhdr}
\usepackage[export]{adjustbox}
\usepackage{graphicx}
\usepackage{amsmath}

\usepackage{hyperref}
\usepackage{verbatim}
\usepackage{mathtools}

\usepackage[backgroundcolor=green!50,linecolor=orange,disable]{todonotes}

\title{Ensemble averaging in JT gravity from entanglement in Matrix Quantum Mechanics }

\author[1]{Gabriele Di Ubaldo,}
\author[2]{Giuseppe Policastro}

\affiliation[1]{Université Paris-Saclay, CNRS, CEA, Institut de Physique Théorique, 91191, Gif-sur-Yvette, France
}

\affiliation[2]{Laboratoire de Physique de l’École Normale Supérieure, ENS, Université PSL, CNRS, Sorbonne Université, Université de Paris, F-75005 Paris, France}

\emailAdd{gabriele.diubaldo@ipht.fr}
\emailAdd{giuseppe.policastro@ens.fr}

\abstract{We consider the generalization of a matrix integral with arbitrary spectral curve $\rho_0(E)$ to a 0+1D theory of matrix quantum mechanics (MQM).   
Using recent techniques for 1D quantum systems at large-$N$, we formulate a hydrodynamical effective  theory for the eigenvalues. The result is a simple 2D free boson BCFT on a curved background, describing the quantum fluctuations of the eigenvalues around $\rho_0(E)$, which is now the large-$N$ limit of the quantum expectation value of the eigenvalue density operator $\hat{\rho}(E)$.\\ The average over the ensemble of random matrices becomes a quantum expectation value.  Equal-time density correlations  reproduce the results (including non-perturbative corrections) of random matrix theory. This suggests an interpretation of  JT gravity as dual to a \textit{one-time-point} reduction of MQM. \\
As an application, we compute the R\'enyi entropy associated to a bipartition of the eigenvalues. 
 We match  a previous result by Hartnoll and Mazenc for the $c=1$ matrix model dual to two-dimensional string theory and extend it to arbitrary $\rho_0(E)$. The hydrodynamical theory provides a clear picture of the emergence of spacetime in two dimensional string theory. The entropy is naturally finite and displays a large amount of short range entanglement, proportional to the microcanonical entropy. We also compute the reduced density matrix for a subset of $n<N$ eigenvalues.
}
\begin{document} 
\maketitle
\flushbottom
\newpage
\input{1introduction}
\input{2hydrormt}

\input{3results}

\input{4Discussion.tex}

\newpage

\bibliography{biblio.bib}{}
\end{document}

%% file: 1introduction.tex
\section{Introduction}
\label{sec:Introduction}

Random Matrix Models have been studied for a long time, as they provide a powerful computational tool and a great source of insight with many applications in different fields, from nuclear physics to condensed matter to high-energy physics.\footnote{  There are many reviews on the subject. For general aspects see, in rough order of complexity \cite{vivo,anninos,Eynard:2015aea,Akemann:2011csh, mehta2004random,Tao}. For applications to low dimensional gravity and string theory see \cite{kazakovreview,klebanov,martinec,ginspargmoore,polchinski,dasjevicki,DiFrancesco:1993cyw,Nakayama:2004vk}. For applications to chaotic systems see \cite{haake,efetov_1996}.}

One of the most intriguing applications arises from the connection to quantum gravity. 
 It was first observed by 't Hooft \cite{tHooft:1973alw} that a theory with matrix degrees of freedom  can be interpreted as a theory of random surfaces in the large-$N$ limit and thus, in many cases, it can be connected to some 2d quantum gravity/string theory.  The perturbative expansion in Feynman diagrams can be reorganized as a topological expansion in the genus of the surface, with $1/N$ playing the role of the expansion parameter (string coupling). This idea has found its most concrete realization so far in the AdS/CFT correspondence \cite{Maldacena:1997re,Witten:1998qj,Gubser:1998bc}. In its most basic and well-understood instance,  this correspondence relates a gravity theory on 5D Anti-de Sitter space to  a $SU(N)$ gauge theory on the 4D boundary.  
Despite the fact that the correspondence has a very precise formulation and has been tested to great accuracy,  its perhaps most striking conceptual aspect, namely the emergence of spacetime from the matrix degrees of freedom,  is still poorly understood. 
The correspondence gives in principle a complete definition of quantum gravity in AdS, since the boundary theory is well defined non-perturbatively (e.g. by the CFT axioms); however, because it is a weak/strong coupling duality, it is still difficult to use it in order to find detailed answers to fundamental questions, such as  the information loss paradox, and the statistical interpretation of the Bekenstein-Hawking entropy in terms of black hole microstates. 
\subsection{Motivation}
Driven by the desire to understand these questions in a simplified setting, there have been many recent developments in low dimensional holography.  The SYK model \cite{Sachdev:1992fk,KitaevTalk} is composed of a large number of fermions interacting with disordered couplings. In the large-N limit the low-energy sector of the model is described holographically by 2-dimensional Jackiw-Teitelboim (JT) gravity, or  equivalently by a 1D  Schwarzian theory \cite{Almheiri:2014cka,Maldacena:2016hyu,Kitaev:2017awl}. 
The SYK model and JT gravity were shown to exhibit quantum chaos as universally described by random matrix theory \cite{Cotler:2016fpe,Saad:2018bqo}.
The  paper \cite{Saad:2019lba} showed a much stronger connection to random matrices: the partition function of JT gravity on a surface of arbitrary genus and number of boundaries agrees with the perturbative expansion of a certain matrix integral, thus solving the theory to all orders in the genus expansion. 
The matrix integral is interpreted as an average over an ensemble of Hamiltonians and the matrix eigenvalues as the energy levels dual to gravitational microstates.  It was also noted that JT gravity can be seen as the $p \to \infty$ limit of  $(2,p)$ minimal strings  which were long known to be  dual to matrix models \cite{Kazakov:1989bc,Staudacher:1989fy,ginspargmoore,DiFrancesco:1993cyw,Seiberg:2004at}.  The study of non-perturbative effects in these models can then help us understand
the detailed structure of gravitational microstates.  As a consequence much effort has been devoted to this pursuit (see \cite{Mertens:2022irh} for a review). 
However many interconnected questions still remain. The matrix integral does not provide a unique non-perturbative completion of JT gravity \cite{Johnson:2022wsr}. The bulk theory does not seem to have, at first sight,  a well defined dual quantum mechanical system, but rather an ensemble of them. The presence of connected geometries and the consequent lack of factorization pose a deep puzzle about the nature of the gravitational path integral \cite{Harlow:2018tqv}. A more explicit understanding  of the emergence of spacetime from the dual degrees of freedom remains to be attained \cite{Das:2017wae}.    \\ 
In this work we discuss some of these issues by considering a generalization of the kind of matrix integral dual to JT-gravity, given by a $0+1D$ theory  of matrix quantum mechanics \cite{planardiags}:
\begin{equation}
S=\int \dd t \, \text{tr}\qty(\frac{1}{2}\dot{H}^2 +V(H)).
\end{equation}
with an arbitrary potential $V(H)$.  
The classical average over the matrix ensemble becomes a quantum path integral:  
\begin{equation}
    \int dH e^{-N\text{tr}V(H)} \rightarrow \int DH(t) e^{-S}.
\end{equation}

We can  think of the original matrix integral as a matrix quantum mechanics with \textit{one-time-point} (as discussed in \cite{Saad:2021rcu} for SYK) meaning that we look at a single instant of time where the dynamics are frozen. We will make this statement precise and show that we can reproduce matrix integral results from equal time correlations in matrix quantum mechanics. 
In particular, we have a quantum density operator $\hat{\rho}(E)$ whose expectation value is the ensemble-averaged density of eigenvalues:
\begin{equation}
    \rho(E)=\braket{\hat{\rho}(E)}
\end{equation}
and similarly for higher correlations.  The spectral curve $\rho_0(E)$, defined as the large $N$ limit of $\rho(E)$, can be chosen to be that of any specific matrix integral.
This offers a possible interpretation  of JT gravity as being dual to a \textit{one-time-point} matrix quantum mechanics with the appropriate spectral curve $\rho^{JT}_0(E)$ \cite{Saad:2019lba}. \\
Matrix quantum mechanics  is richer than a a matrix integral, first and foremost because it is a quantum mechanical theory. The eigenvalues are described by a wavefunction $\psi_N(E_1,\dots,E_N)$ instead of a classical probability distribution $\rho_N(E_1,\dots,E_N)$ as in random matrix theory. Since the matrix eigenvalues describe the microstates $\{E_i\}$ of JT gravity, we might think of matrix quantum mechanics as describing their associated wavefunctions  $\ket{E_i}$. It is then natural to consider the entanglement between eigenvalues. The average over the ensemble of random matrices becomes a quantum expectation value  in Hilbert space. Thus, the statistical fluctuations due to ensemble averaging over Hamiltonians  may now be interpreted as quantum fluctuations of a single quantum mechanical system.  These observations point to Matrix Quantum Mechanics as an interesting  generalization of the matrix integral dual to JT gravity.  \\ 
 In two spacetime dimensions there is another instance of holographic duality: the duality between two-dimensional string theory and the $c=1$ matrix model \cite{kazakovmigdal}, a theory of matrix quantum mechanics with a specific potential. This duality precedes AdS/CFT and has been extensively checked both perturbatively and non-perturbatively. \footnote{ See \cite{martinec,klebanov,ginspargmoore,polchinski} for reviews. See  \cite{Balthazar:2017mxh,Balthazar:2018qdv,Balthazar:2019rnh,Balthazar:2019ypi,Balthazar:2022atu,Balthazar:2022apu} for extensive recent work on matching scattering amplitudes. See  \cite{Karczmarek:2004bw,Betzios:2017yms,Betzios:2016yaq,Betzios:2022pji,Ahmadain:2022gfw} for recent related work on black holes}
Two-dimensional string theory and JT gravity form part of the same family of theories.  A minimal string consists of a Liouville CFT with $c_L>25$ and a minimal model with $c_M<1$ coupled by anomaly cancellation. In the limit $p\rightarrow \infty $ of the $(2,p)$ minimal string, which corresponds to JT gravity, we have that $c_M\rightarrow -\infty$. Instead, two dimensional string theory consists of a $c_L=25$ Liouville theory and a $c_M=1$ free boson. Thus JT gravity  and two-dimensional string theory  lie at opposite ends of the same spectrum of worldsheet theories given by Liouville theory coupled respectively to $c_M=-\infty$ and $c_M=1$.\footnote{For a review of the Liouville approach, see \cite{Nakayama:2004vk}}
Despite knowing they are related, the  relation between the two dualities has yet to be understood explicitly (See \cite{Mertens:2020hbs,Banks:2020zrt,Betzios:2020nry,Suzuki:2021zbe} for related work). From the matrix model point of view, JT gravity is dual to a matrix integral over a single matrix  while, by discretizing time, matrix quantum mechanics can be thought of as the continuum limit of a chain of $q$ matrices \cite{BANKS1990279}. Thus JT is dual to a single matrix while two-dimensional string theory is dual to an infinite number of matrices, one for each instant of time.
Understanding better the relationship between the two dualities could help elucidate various aspects of JT gravity.
For example, in 2D string theory the dual is a single quantum mechanical system and no averaging is involved. Spacetime can be thought of as emergent from the continuum of eigenvalues at large $N$ and  locality can be probed by the entanglement between the eigenvalues \cite{Hartnoll:2015fca, Das:1995jw,Das:1995vj}, as we will demonstrate. The worldsheet description present in minimal and 2D string theories allows for a detailed study of non-perturbative effects  \cite{Balthazar:2019rnh,Balthazar:2019ypi,Balthazar:2022apu,Sen:2020eck,Sen:2020oqr,Sen:2019qqg,Eniceicu:2022xvk,Chakravarty:2022cgj,Eniceicu:2022dru,Sen:2022clw,Eniceicu:2022nay} and their matching to the matrix model.
Understanding the relation between JT gravity and two-dimensional string theory at the level of the dual matrix models motivates a new consideration of matrix quantum mechanics.\\
Finally, the duality between the $c=1$ matrix model and  two-dimensional string theory is a perfect playground to study the emergence of spacetime from matrix degrees of freedom in gauge theories since, at large $N$, the eigenvalues form a continuum that is directly related to the dual spacetime.  
The relation between spacetime and eigenvalue-space can be tested in various ways, \textit{e.g.} using local observables, scattering of the excitations,  or using  entanglement entropy, as was done in \cite{Hartnoll:2015fca}. The motivation of this last paper was to apply to the $c=1$ matrix model the insight, gained in AdS/CFT with the Ryu-Takayanagi formula, of the essential role that entanglement plays in the emergence of spacetime \cite{Ryu:2006bv,VanRaamsdonk:2010pw}. In this paper we will give a different and more comprehensive perspective on the  eigenvalue/spacetime relation by explicitly constructing the geometry of eigenvalue-space that corresponds to the spacetime geometry in a natural way and studying its entanglement properties. The entanglement between eigenvalues is an example of entanglement in target space. Characterizing the entanglement of target space degrees of freedom is essential to understand spacetime in string theory and holography, and recently there has been a growing interest in the subject, see \cite{Mazenc:2019ety,Das:2020xoa,Das:2020jhy,Gautam:2022akq,Sugishita:2021vih,Tsuchiya:2022ffu,Frenkel:2021yql,Das:2022mtb} .  \\

\subsection{Overview and results}
We start sec. \ref{sec:model} by recalling some basic facts about Matrix Quantum Mechanics. Eigenvalue repulsion enforces fermionic statistics for the eigenvalues which can be mapped to a system of  fermions in an external potential.  We introduce a second-quantized fermionic field $\Psi(E)$ which gives the eigenvalue density operator $\hat{\rho}(E)=\Psi^{\dagger}(E)\Psi(E)$. The density of eigenvalues is the expectation value $\rho(E)=\braket{\hat{\rho}(E)}$ which is, at leading order in the large $N$ limit, equal to the spectral curve $\rho(E)\approx \rho_0(E)$.
We then proceed in sec. \ref{ssec:hydro} to illustrate the construction of an effective  hydrodynamical theory for the eigenvalues  valid for arbitrary $\rho_0(E)$ . The construction follows   from recent developments in the study of 1D many body quantum systems in external potentials \cite{dubailcalab,introhydro1,introhydro2}. It can be seen as a generalization of   the collective field theory approach \cite{dasjevicki,JEVICKI1980511} to arbitrary potentials. 
In sec. \ref{ssec:qhydro} we discuss quantum fluctuations of the eigenvalues in  the effective theory. One can show that the quantum hydrodynamical fluctuations of the eigenvalues are described by a 2D free boson CFT on a curved background determined by $\rho_0(E)$ with boundaries at the edge of the spectrum where $\rho_0(E^*)=0$. 

In section \ref{sec:CFT} we proceed to use the 2D CFT to study the different properties of the eigenvalues. We start by computing spectral correlations in sec. \ref{ssec:correlations} which are now given by correlation functions of the density operator: $\braket{\hat{\rho}(E)}$ and $\braket{\rho(E_1)\rho(E_2)}$. These are given by correlation functions of vertex operators in the CFT. We reproduce the leading non-perturbative corrections to the density of states $\rho(E)$ and to the level-correlation $\rho(E_1,E_2)$ as described in sec. 5 of \cite{Saad:2019lba} by considering equal-time correlations. In other words, we reproduce the oscillations of $\rho(E)$ around the semiclassical density $\rho_0(E)$ and the terms in $\rho(E_1,E_2)$ corresponding to the ramp and plateau in the spectral form factor (i.e. the sine kernel).  In matrix quantum mechanics these spectral correlations arise due to quantum fluctuations of a single quantum mechanical system, as opposed to statistical fluctuations due to ensemble averaging. 
This matching provides evidence to support the idea that a matrix integral and consequently JT gravity might be interpreted as a \textit{one-time-point} matrix quantum mechanics with the same spectral curve $\rho_0(E)$. 

 In sec. \ref{ssec:entent} we consider the entanglement between the eigenvalues. We compute the R\'enyi entropies for a bipartition $(0,E)\cup (E,E_R)$, where $E_R$ is the right edge of the eigenvalue density, finding some interesting features.  For non double-scaled matrix models, where the density has a right edge $E_R$, we see that the entanglement entropy follows a ``Page curve'' (as a function of the lenght of the interval) as required by unitarity and comes down instead of growing indefinitely. This feature is lost in double-scaled models where $E_R\rightarrow \infty$ indicating that indeed we are missing states from the spectrum. The entanglement entropy is naturally finite due to the mean spacing between the eigenvalues $\frac{1}{\rho_0(E)}\sim e^{-S_0}$ acting as a UV cutoff.\footnote{While finishing  this paper, the work \cite{Das:2022nxo} appeared which discusses the finiteness of the entanglement entropy in matrix quantum mechanics. Their methods are different and the discussion is complementary. In particular, they discuss a vanishing potential $V=0$  while we treat arbitrary potentials.} In two-dimensional string theory we can interpret the finiteness of the entropy as due to $g_s$ stringy effects regulating the divergence as first noted in \cite{Hartnoll:2015fca}. 
 We notice that the entanglement entropy $S_{ent}(E)$ present a leading contribution proportional to the microcanical entropy in the window $E\pm dE$ such that $S_{ent}(E)\propto S_0(E)=\log(\rho_0(E))$, indicating a large amount of short range entanglement between eigenvalues close to the boundary.  We also compute the entanglement entropy for an interval bipartition $(E_1,E_2)$, extending the results of \cite{Hartnoll:2015fca} for the $c=1$ matrix model to arbitrary spectral curves $\rho_0(E)$.
 We provide constructive evidence for the proposed map between the eigenvalue-space and the emergent spacetime in two-dimensional string theory\cite{Das:1995vj,Hartnoll:2015fca} and the identification of the spacetime geometry with the geometry of the Fermi surface. 
 
In sec. \ref{ssec:dm} we compute the one eigenvalue reduced density matrix obtained by tracing out $N-1$  eigenvalues, corresponding to the fermion one-body density matrix $g(E,E')=\braket{\Psi^{\dagger}(E)\Psi(E')}$. We also write the general expression for the $n$ eigenvalue density matrix. 

We conclude in sec. \ref{sec:discussion} with a discussion of open questions and possible future work.

%% file: 2hydrormt.tex
\section{Quantum hydrodynamics of random matrix eigenvalues}
\label{sec:model}
We study the quantum mechanics of a random $N\times N$ hermitian matrix $H(t)$ in a generic potential $V(H)$ with the following action:  
\begin{equation}
S= N\int \dd t \, \text{tr}\bigg[\frac{1}{2}\dot{H}^2 +V(H)\bigg].
\end{equation}

The eigenvalues $(E_1\dots E_N)$ of $H$ no longer obey a classical probability distribution as in Random Matrix Theory.  Instead they are now described by a  quantum mechanical wavefunction $\psi_N(E)$. We will now briefly summarize some well known facts about matrix quantum mechanics (MQM) and derive the Schrodinger equation for the $N$-eigenvalue wavefunction $\psi_N(E)$. More details can be found  in the above mentioned reviews \cite{kazakovreview, klebanov,polchinski,martinec, anninos,ginspargmoore}.

\subsection{Eigenvalues as fermions}
To study the eigenvalues we  diagonalize the matrix $H$: 
\begin{equation}
H=\Omega^{\dagger}E \, \Omega
\end{equation}
where $\Omega \in SU(N)$ and $E=\text{diag}(E_1,\dots, E_N)$.
This change of variables has a non-trivial jacobian which modifies the path integral measure $\mathcal{D}H(t)$:
\begin{equation}
   \int \mathcal{D}H=\int \mathcal{D}\Omega \prod_i \mathcal{D}E_i \Delta^2(E),
\end{equation}
where $\Delta(E)=\prod_{i<j}(E_i-E_j)$ is the well known Vandermonde determinant which causes eigenvalue repulsion in random matrix theory. We will now see that in MQM, eigenvalue repulsion becomes the Pauli exclusion principle resulting in fermionic eigenvalues \cite{planardiags}. 
Due to the non-trivial Jacobian, the kinetic term for the eigenvalues becomes: 
\begin{equation}
-\frac{1}{2}\sum_{i=1}^{N}\frac{1}{\Delta^2(E)}\dv{E_i}\Delta^2(E)\dv{E_i}.
\end{equation}
Thanks to the fact that $\sum_i \dv[2]{\Delta}{E_i}=0$, this is equal to:
\begin{equation}
    -\frac{1}{2\Delta}\sum_i\dv[2]{E_i}\Delta.
\end{equation}
The Hamiltonian $\mathcal{H}$ of matrix quantum mechanics, after diagonalization of $H$,  is then \cite{planardiags,klebanov}:
\begin{equation}\label{eq:hamiltonianmqm}
    \mathcal{H}=  -\frac{1}{2\Delta}\sum_i\dv[2]{E_i}\Delta +\sum_i V(E_i) +\sum_{i<j}\frac{L_{ij}^2+\tilde{L}_{ij}^2}{(E_i-E_j)^2}. 
\end{equation}
The first term is the kinetic term for the eigenvalues we just discussed. The matrix potential $V(H)$ becomes a single particle potential for the eigenvalues $V(E_i)$ due to invariance of the trace. The last term is the kinetic term for the angular degrees of freedom $\Omega$, where  $L_{ij},\tilde{L}_{ij}$ are the angular momenta and $(E_i-E_j)^2$ plays the role of a radius in the direction $ij$.
Let us denote a generic wavefunction for the Hamiltonian $\mathcal{H}$  as $\chi_N(E,\Omega)$, which will depend on both the eigenvalues $E$ and the angular variables $\Omega$. We use the subscript $N$ as a reminder that the wavefunctions depend on all the eigenvalues $E_1,\dots, E_N$.
We will restrict ourselves to scalar configurations which are invariant under $SU(N)$ rotations, namely the singlet sector. Thus we consider  wavefunctions $\chi_N(E)$ which are independent of the angular variables. The singlet wavefunctions $\chi_N(E)$ are the relevant ones and correctly describe MQM in the following regimes:\\

$\bullet$ \textbf{Ground state.} Since the angular  term is positive definite, the ground state of the system is given by the singlet sector ground state . Thus the singlet wavefunction describes the collective ground state of the $N$ eigenvalues. \\

$\bullet$ \textbf{Low temperature phase.}
Considering MQM at finite temperature, we have a Berezinski-Kosterlitz-Thouless transition at $\beta_{BKT}$. The singlet sector describes the low temperature phase  $\beta>\beta_{BKT}$  \cite{Gross:1990md,Gross:1990ub,Yang:1990ns,Boulatov:1991xz}.\\

Consider now the Schrodinger equation for the singlet sector $\mathcal{H}\chi_N(E)=\epsilon \chi_N(E)$. The wavefunctions $\chi_N(E)$ are clearly symmetric functions of the eigenvalues. By defining a completely anti-symmetric wavefunction $\psi_N(E)=\Delta(E)\chi_N(E)$, the Schrodinger equation now reads:
\begin{equation}
   \sum_{i=1}^N \mathcal{H}_i\psi_N(E)=\sum_i^N \epsilon_i \psi_N(E), \qquad \mathcal{H}_i=-\frac{1}{2}\dv[2]{E_i}+V(E_i).
\end{equation}
The Hamiltonian acting on $\psi_N(E)$ is now a sum of single-particle Hamiltonians. The wavefunction $\psi_N(E)$ is completely antisymmetric by construction due to the antisymmetry of the Vandermonde and  vanishes whenever $E_i=E_j$. We see that the eigenvalue repulsion of random matrices  enforces the Pauli exclusion principle. The eigenvalues $E_i$ are then equivalent to a system of $N$ fermions each in an external potential $V(E)$, interacting only through the exclusion principle/eigenvalue repulsion. \\
The many-body ground state wavefunction $\psi_N(E)$ can be obtained  by first solving for the single particle wavefunctions $\psi_{\epsilon}(E)$ and building the Slater determinant $\psi_N(E)=\frac{1}{\sqrt{N!}}\text{det}_{ij}(\psi_{\epsilon_i}(E_j))$ which involves a single fermion in each energy level up to the Fermi energy $\epsilon_F$, the energy of the last ($N$-th) fermion. We will not do this as it involves solving the Schrodinger equation for a specific choice of potential with the resulting Slater wavefunctions $\psi_N$ being complicated expressions for large $N$.
We will instead describe the system in second quantization  by introducing a fermionic field $\Psi(E)$ \cite{moore,Hartnoll:2015fca} with the following Hamiltonian $\mathbb{H}$ :
\begin{equation}
    \mathbb{H}= N \int dE \Psi^{\dagger}(E)\qty(-\frac{1}{2 N^2}\dv[2]{E}+V(E))\Psi(E).
\end{equation}

 The fermionic field $\Psi(E)$ can be expressed as a mode expansion with creation/annihilation operators $a_{\epsilon},a^{\dagger}_{\epsilon}$ weighted by the single particle wavefunctions $\psi_{\epsilon}(E)$:
 \begin{equation}
     \Psi(t,E)=\int d\epsilon e^{-i\epsilon t}a_{\epsilon}\psi_{\epsilon}(E).
 \end{equation}
The fermions  fill the potential $V(E)$ up to the Fermi energy $\epsilon_F$. We can control how  the potential is filled by introducing a chemical potential $\mu= N \epsilon_F$ in the Hamiltonian $\mathbb{H}-\mu\Psi^{\dagger}\Psi$. The system forms a Fermi surface $\ket{\mu}$ on which the operators $a_{\epsilon},a^{\dagger}_{\epsilon}$ act as follows:
\begin{equation}
\begin{split}
    &a_{\epsilon}\ket{\mu}=0 \quad \epsilon >\mu, \\
    & a^{\dagger}_{\epsilon}\ket{\mu}=0 \quad \epsilon <\mu.
\end{split}
\end{equation}\\
The presence of a Fermi sea  corresponds to having a finite density of eigenvalues $\rho_0(E)$ in RMT.
 In what follows we will employ  recent techniques from condensed matter \cite{Dubail:2016tsc}  describing the quantum fluctuations of the Fermi surface by a 2D effective hydrodynamical theory .\\
 Correspondingly, one can develop a quantum hydrodynamical theory for the eigenvalue density $\rho(E)$, describing the fluctuations around a semiclassical background $\rho_0(E)$ given by the RMT spectral curve.  Quantum fluctuations on top of the Fermi surface involving the creation/annihilation of a single eigenvalue will produce non-perturbative effects in $\frac{1}{\rho_0(E)}\sim e^{-S_0}$ (similarly as described in sec. 5 of \cite{Saad:2019lba}).
 The effective theory will be a simple free boson CFT on a curved background with a boundary. 
This simple description  allows us to study many interesting features of Matrix Quantum Mechanics. We can access non-perturbative physics like the oscillations in the density of states $\rho(E)$ and the plateau in the two level correlation $\rho(E_1,E_2)$ which are a consequence of the underlying discreetness of the spectrum of $H$. 
 We can also compute observables that do not have a clear classical counterpart such as the reduced density matrix obtained by tracing out $k$-out-of-$N$ eigenvalues and the spectrum of Renyi entropies for arbitrary bipartition $(E_1,E_2)$. 
 \\
 Incorporating the chemical potential, we arrive  at the following Hamiltonian, which is the starting point for the rest of discussion:
\begin{equation}\label{eq:fermionFT}
     \mathbb{H}= \int dE \Psi^{\dagger}(t,E)\qty(-\frac{1}{2}\dv[2]{E}+(V(E)-\mu))\Psi(t,E).
\end{equation}

\subsection{Effective hydrodynamics of the eigenvalue density $\rho(E)$}\label{ssec:hydro}
We now give a self-contained review of some recent developments in the study of 1D many-body quantum systems in external potentials via hydrodynamics \cite{dubail, dubailcalab, brundubaildm, brundubailgff}.
The hydrodynamics approach to 1D quantum systems was introduced a few years ago in \cite{introhydro1,introhydro2} and has been rapidly developing ever since, see the recent lectures \cite{doyon} for a review.  We will only introduce the necessary tools for our purposes. \\
Conformal Field Theory in 2D is a well-proven  technique in addressing 1D critical quantum systems \cite{giamarchi}.   It is commonly used to describe low energy excitations around a fixed energy scale (such as the Fermi energy $\epsilon_F$) and so it is not a priori possible to apply it to inhomogeneous systems, where there is a varying energy scale due  to an external potential or out-of-equilibrium dynamics. 
On the other hand, hydrodynamics is useful to describe inhomogeneous systems at mesoscopic scales, large enough to contain a macroscopic number of degrees of freedom but smaller than the characteristic scale of the inhomogeneities.  In  \cite{dubailcalab},  they obtained a 2D CFT describing inhomogeneous 1D quantum systems using hydrodynamics. The CFT lives on a non-trivial background metric encoding the inhomogeneities of the original system. \\
Let us start by considering a many-body quantum system composed of $N$ particles with a finite density  $\rho(x)$ in the large $N$ limit in an interval $x\in (x_L,x_R)$. This means that the quantum density operator $\hat{\rho}(x)=\Psi^{\dagger}(x)\Psi(x)$ acquires a VEV $\rho(x)\equiv \braket{\Psi^{\dagger}(x)\Psi(x)}$. The VEV introduces a length scale in the system corresponding to the local average spacing between particles $d(x)=\frac{1}{\rho(x)}$.  In Random Matrix Theory there is a finite density of eigenvalues $\rho(E)$ due to eigenvalue repulsion, which is analogous to the non-zero VEV of the quantum density operator $\hat{\rho}$. We will make this correspondence precise in MQM: since the eigenvalues are fermions we have $x=E$ and we have that $\rho(E)=\braket{\hat{\rho}(E)}$. The mean level spacing is then equal to $d=\frac{1}{\rho(E)}\sim e^{-S_0}$.\footnote{ This is not the first instance where the spectral density  $\rho(E)$ is identified with a VEV, in \cite{Altland:2020ccq}\\ $\rho(E)$ is identified as the order parameter responsible for Causal Symmetry Breaking in the universal late time behaviour of chaotic systems.} 
The key assumption to develop hydrodynamics for inhomogeneous systems is the separation of scales, meaning there exists an intermediate mesoscopic scale $\ell$ such that:
\begin{equation}\label{eq:cond1}
    d\ll \ell \ll \frac{\rho(x)}{\partial_x\rho(x)},
\end{equation}
where  $\frac{\rho}{\partial \rho}$ is the characteristic length scale of the inhomogeneities. 
The scale $\ell$ is then small enough such that the system is quasi-homogeneous and large enough to contain a thermodynamically large number of particles. 
These scales provide both UV and IR cutoffs in the effective theory defined at energy scales  $\Lambda$ such that: 
\begin{equation}\label{eq:cond1en}
  \frac{\partial_x\rho(x)}{\rho(x)} = \Lambda_{IR}\ll \Lambda \ll \Lambda_{UV}=\frac{1}{d}=\rho(x),
\end{equation}
Having understood the characteristic scales and the regime of validity of the effective theory, we will now focus on a specific system:  the Lieb-Liniger gas of interacting bosons in an external potential. It is defined by the following Hamiltonian:
\begin{equation}
    H=\int dx \qty(\Phi^{\dagger}\qty(\frac{\hbar^2\partial_x^2}{2m} +V(x))\Phi +\frac{g}{2}{\Phi^{\dagger}}^2\Phi^2),
\end{equation}
where $\Phi(x)$ is a bosonic field $[\Phi(x),\Phi(x')]=\delta(x-x')$.
This system can be solved exactly via Bethe-Ansatz in the homogeneous $V=0$ case \cite{korepin}.
In the limit of hard-core bosons $g\rightarrow \infty$ it is equivalent to a system of free fermions in the potential $V(x)$ and thus describes the eigenvalues of MQM.  This limit is often referred to as the Tonks-Girardeau gas in the literature. 
The mapping between  hard-core bosons and free fermions is made via a Jordan-Wigner string:
\begin{equation}
    \Psi^{\dagger}(x)=e^{i\pi \int_{y<x}\Phi^{\dagger}(y)\Phi(y)dy }\Phi^{\dagger}(x).
\end{equation}
$\Psi(x)$ is now a fermionic field $\{ \Psi^{\dagger}(x),\Psi(x')\}=\delta(x-x')$. We  then obtain the free fermion hamiltonian:
\begin{equation}
    \mathbb{H}=\int dx \Psi^{\dagger}\qty(-\frac{\hbar^2\partial_x^2}{2m}+V(x))\Psi.
\end{equation}

The hydrodynamic description of  the homogeneous case ($V=0$) was first presented in \cite{integrablehydro} where the authors developed the hydrodynamics of out-of-equilibrium systems with an infinite number of conserved charges.
Let us now proceed with the case of a general external potential $V(x)$. We start by writing down the Euler equations for a Galilean invariant fluid in the presence of an external potential: 
\begin{equation}
\begin{split}
       \partial_t \rho +\partial_xj&=0, \\
       \partial_tu+u\partial_xu+\frac{1}{\rho}\partial_xP&=-\partial_xV,
\end{split}
\end{equation}
where $\rho$ is the particle density,  $u$ is the mean velocity given by $u=\frac{j}{\rho}$ with $j$ the momentum density and $P$ is the pressure. To close the system of equations we need the equation of state at zero temperature which expresses the pressure as a function of the density $P(\rho)$. This can be obtained from the energy density $\rho_E$ by the thermodynamic relation $P(\rho)=-\rho_E+\rho\qty(\frac{\partial \rho_E}{\partial\rho})_{T=0}$.
In the Lieb-Liniger model,  these equations follow from the conservation of the following charges: 

\begin{equation}
    \begin{split}
        \hat{\rho}(x)&=\Phi^{\dagger}(x)\Phi(x)\\
        \hat{\rho}_P(x)&=-i\hbar \Phi^{\dagger}(x)\partial_x\Phi(x)\\
        \hat{\rho}_E(x)&=\frac{\hbar^2}{2}\qty(\partial_x\Phi^{\dagger}(x)\partial_x\Phi(x))+\frac{g}{2}\Phi^{\dagger}{}^2(x)\Phi^2(x)
    \end{split}
\end{equation}
with associated currents $\hat{j},\hat{j_P},\hat{j_E}$.
The Euler equations describe the  expectation values of the charges and currents $\braket{\rho}=\rho,\braket{j}=j$.\footnote{Since we will be considering zero temperature and thus zero entropy hydrodynamics, the continuity equation for the energy density is trivially satisfied and we have not displayed it in the text. }
 In particular, consider the quantum expectation value of the density operator for the fermionic field $\Psi$ : 
\begin{equation}
    \rho(x)\equiv\braket{\Psi^{\dagger}(x)\Psi(x)}.
\end{equation}
In the hydrodynamic description where the system is quasi-homogeneous at the scales we are probing, we will have a semiclassical background density $\rho_0(x)$ at leading order in  the large $N$ limit such that $\rho(x)\approx \rho_0(x)$. We can then describe fluctuations of this semiclassical density which will give both subleading corrections to $\rho(x)$ and correlations $\braket{\hat{\rho}(x)\hat{\rho}(x')}$. The semiclassical density sets the scales for which the effective theory is valid and self-consistent:
\begin{equation}\label{eq:cond}
    \frac{\partial_x\rho_0(x)}{\rho_0(x)} \ll \Lambda \ll \rho_0(x).
\end{equation}
In Random Matrix Theory the density $\rho_0(x)$ is given by the leading density of eigenvalues $\rho_0(E)$ in the large $N$ limit, meaning the spectral curve/the disk density of states.\\
We can obtain an approximate expression for $\rho_0(x)$ as a function of the potential $V(x)$ in the hydrodynamical effective theory. Since $\rho_0(x)$ is the density at equilibrium, meaning $\partial_t\rho=0, u=0$,  the Euler equation reduces to $\frac{1}{\rho}\partial_xP=-\frac{1}{m}\partial_xV$. Using the thermodynamic relation $dP=\rho SdT+\frac{\rho}{m}d\mu$ at $T=0$ we have that $\partial_x(\mu(x)+V(x))=0$.  The local chemical potential is then $\mu(x)=\mu-V(x)$ where $\mu$ is the fixed chemical potential appearing in the Hamiltonian.  For a homogeneous system, the  equilibrium  density  is just  a function  of the chemical potential $\rho_{hom}=\rho_{hom}(\mu)$ so for scales where the system is quasi-homogeneous we can simply substitute the local chemical potential $\mu(x)$ in the homogeneous density. We have then  that the semiclassical density $\rho_0(x)$ in the hydrodynamic description is given by:
\begin{equation}
    \rho_0(x)\overset{hydro}{\approx} \rho_{hom}(\mu(x)), \qquad \mu(x)=\mu-V(x).
\end{equation}
The theory will be entirely defined in terms of the density $\rho_0(x)$, without making reference to the potential so it is not necessary to use the relation above although it can be useful if we wish to define our MQM by the potential $V(x)$ instead of the spectral curve. \\
In particular, the density for free fermions with $V(x)=0$ is:
\begin{equation}
    \rho_{hom}(\mu)=\frac{\sqrt{2\mu}}{\pi \hbar},
\end{equation}
we then have that the semiclassical density is:
\begin{equation}
    \rho_0(x)\approx \frac{1}{\pi \hbar}\sqrt{2(\mu-V(x))}.
\end{equation}
This expression is usually called the Local Density Approximation (LDA) and it is well-known that it correctly describes the bulk density (sufficently away from edges where $\rho\approx 0$) of a Fermi gas in the large $N$ limit \cite{castin,Dean_2016}.
We can see immediately that for a Gaussian potential $V(x)=\frac{x^2}{2}$ it correctly reproduces Wigner's semicircle law.
We can also see that this is exactly the expression for the momentum $p(x)$ of a particle with energy $\mu$ appearing in the WKB approximation:
\begin{equation}
    \psi_{WKB}\approx  \frac{A}{\sqrt{p(x)}}\exp(\pm \frac{i}{\hbar}\int^x p(x')), \qquad p(x)=\pi \hbar \rho_0(x).
\end{equation}

As a final consistency check note that the density of free fermions scales as $\rho\sim \mathcal{O}(\hbar^{-1})$ so that the mean spacing $d=\frac{1}{\rho}\sim \mathcal{O}(\hbar)$ while the scale of inhomogeneities  is $\frac{\rho}{\partial_x\rho}\mathcal{O}(1)$.
Thus for $\hbar \rightarrow 0$ there is indeed separation of scales and we can always find a regime  $\ell\sim \mathcal{O}(\hbar^{\nu})$ with $0<\nu<1$ where the hydrodynamic description is valid. 
If we send $\hbar \rightarrow 0$, the total number of particles $N=\int \rho(x) dx$ diverges so we should take the large $N$ limit with $N\hbar=\mathcal{O}(1)$. This is the well known property that large $N$ limits are semiclassical. 

\subsection{2D CFT for the quantum fluctuations of $\rho(E)$}\label{ssec:qhydro}
We are now ready to build the field theory for the hydrodynamical fluctuations of the density $\rho(x)$. 
Let us now consider a microscopic correlation function of local operators  $O(x)$ in the ground state:
\begin{equation}
     \braket{O(x_1)\dots O(x_n)}\equiv \lim_{\beta\rightarrow \infty}\frac{\text{tr}\qty[O(x_1)\dots O(x_n)e^{-\beta \mathbb{H}}]}{\text{tr}e^{-\beta \mathbb{H}}},
\end{equation}
where $\mathbb{H}$ is the fermion Hamiltonian in eq. \ref{eq:fermionFT}.

In the hydrodynamic limit where $\frac{1}{N}\sim \hbar \rightarrow 0$ we can compute the correlation function by doing a path integral over the hydrodynamic fields $\rho(x,\tau)$ and $j(x,\tau)$ with an Euclidean action $S_E[\rho,j]$: 
\begin{equation}
    \braket{O(x_1)\dots O(x_n)}\overset{\hbar,\frac{1}{N}\rightarrow 0}{=}\frac{1}{Z}\int D\rho Dj \delta(\partial_{\tau}\rho+\partial_x j)O(x_1)\dots O(x_n) e^{-S_E[\rho,j]},
\end{equation}
where the continuity equation is a constraint in the space of configurations $(\rho,j)$ of the path integral\footnote{The same type of path integral has also appeared recently in \cite{Doyon:2022rkm} as the action for ballistic MFT, although in that case it is supposed to apply at finite temperature and describe statistical fluctuations (thanks to T. Yoshimura for pointing it out).}. 
The task now is to determine the action $S_E[\rho,j]$ that computes these correlation functions in the hydrodynamic limit. 
To do so we will proceed  by first finding an action which gives the Euler equations as its equations of motion. 
We consider the following action:
\begin{equation}\label{eq:hydro}
    S[\rho,j]=\int dx dt \qty(\frac{j^2}{2\rho}+\rho_E(\rho)+(V(x)-\mu)\rho  ).
\end{equation}

We now perform a variation of the action $(\bar{\rho}+\delta\rho,\bar{j}+\delta j)$ starting from  a configuration $(\bar{\rho},\bar{j})$ which satisfies the Euler and continuity equations. 
To perform a variation consistent with the constraint $\partial_t\rho+\partial_x j=0$ we write the fluctuations as:
\begin{equation}
    \delta \rho(x,t)=\frac{1}{2\pi}\partial_xh(x,t), \quad \ \delta j(x,t)=-\frac{1}{2\pi}\partial_th(x,t).
\end{equation}

We have introduced a new field $h(x,t)$ such that the constraint is now trivially satisfied due to the fact that partial derivatives commute.
The second order variation $\delta^2S_E[\bar{\rho}+\delta\rho,\bar{j}+\delta j]$ gives a  quadratic action for the quantum fluctuations described by the field $h(x,t)$. The action is the following:
\begin{equation}\label{eq:fluct}
    S[h]=\frac{1}{8\pi}\int \frac{\sqrt{g}\dd[2]{x}}{K(x)}g^{ab}\partial_a h\partial_b h,
\end{equation}
where $K$, known as the Luttinger parameter, is a function of the  density $\bar{\rho}(x)$
\begin{equation}
    K(x)=\frac{\pi \hbar \bar{\rho}(x)}{v(x)} \quad \text{with} \quad v(x)=\sqrt{\bar{\rho}(x) \partial^2_{\rho} \rho_E},
\end{equation}
and the metric is given by
\begin{equation}
    ds^2=(dx-(u+v)dt)(dx-(u-v)dt).
\end{equation}
where $u(x)=\frac{\bar{j}}{\bar{\rho}}$ is the background local fluid velocity. We see from the metric that $v(x)$ is the local speed of sound in the fluid since excitations  which propagate along light-rays correspond to sound waves propagating at velocity $u\pm v$. The system exhibits curved lightcones due to the  dependence on the local value of the density $\bar{\rho}(x)$\cite{Dubail_2017}.
This metric is the effective geometry of the Fermi surface  seen by the excitations. \\
The above action thus describes quantum fluctuations around non-trivial hydrodynamical backgrounds $\bar{\rho}(x,t),\bar{j}(x,t)$ for 1D inhomogenous quantum systems specified by their microscopic equation of state $\rho_E(\rho)$, from which we obtain the Luttinger parameter $K$ and the sound velocity $v$. \\
There can be corrections to the effective action for the fluctuations \ref{eq:fluct} by expanding to higher order the hydrodynamic action \ref{eq:hydro}. There can also be hydrodynamic gradient corrections, recently discussed in \cite{Riggio_2022}. 
\\
We will now restrict ourselves to the case of free fermions  since we wish to describe  fluctuations of the eigenvalues of random matrices. We also restrict ourselves to equilibrium configurations given by the saddle point $(\bar{\rho},\bar{j})=(\rho_0(x),0)$ where $\rho_0(x)$ is the semiclassical density. We leave the study of out-of-equilibrium dynamics of the eigenvalues for future work. 
 The equation of state for free fermions at zero temperature is:
\begin{equation}
    \rho_E= \frac{\pi^2\hbar^2\rho^3}{6}.
\end{equation}
We have that the sound velocity is simply  proportional to the density
\begin{equation}
    v(x)=\pi\hbar \rho_0(x),
\end{equation}
and the Luttinger parameter  is simply $K=1$.
Notice that $v(x)$ is equal to the classical momentum $p(x)$ appearing in the WKB approximation.
We arrive then at the following Euclidean action, describing the quantum hydrodynamical fluctuations  of free fermions in an external potential: 
\begin{equation}\label{eq:hydroaction}
    S[h]=\frac{1}{8\pi}\int \sqrt{g}\dd[2]{x} g^{ab}\partial_a h \partial_b h ,
\end{equation}
with metric given by (now in units where $\hbar=1$): 
\begin{equation} \label{eq:metric}
    ds^2=\pi^2\rho_0^2(x)d\tau^2+dx^2,
\end{equation}

The action $S[h]$ provides a description of the eigenvalue fluctuations around the semiclassical  spectral density $\rho_0(E)$ in terms of quantum hydrodynamics of the Fermi surface. 
Analogously to the $\frac{1}{N}$ genus expansion of matrix models, which is completely fixed by Topological Recursion in terms of the spectral curve $\rho_0(E)$ (see \cite{Eynard:2015aea} for an extensive explanation),  the theory of hydrodynamic fluctuations of the eigenvalues is  completely determined by the matrix model spectral curve $\rho_0(E)$. The theory is defined on the domain $(x,\tau)\in (x_L,x_R)\times \mathbb{R}$, where $(x_L,x_R)$ are the points where the semiclassical density vanishes $\rho_0(x_{L,R})=0$\footnote{We are treating the case of a single interval with non-zero density, corresponding to single cut matrix models }. 
As long as we work at scales $\Lambda$ inside the range of validity of the hydrodynamic effective theory given in eq. \ref{eq:cond},  we can apply the theory  to a matrix model specified by its semiclassical density of eigenvalues  $\rho_0(E)$. 
In particular, to apply this description to JT gravity, $(2,p)$ minimal strings and related models it is enough to use the spectral density of the desired model, i.e.  $\rho(E)=\rho_{JT}(E)$.
On the other hand, if one wishes to specify the matrix model potential $V(E)$, we can obtain the spectral density in the hydrodynamic approximation as:
\begin{equation}\label{eq:lda}
    \rho_0(E)=\sqrt{2(\mu-V(E))}. 
\end{equation}
We will use this expression for the density to study the $c=1$ matrix model which corrresponds to an inverted oscillator potential $V(E)=-\frac{E^2}{2}$.

The hydrodynamic theory is valid in the domain of non-vanishing  particle density so the theory has a boundary at points $x_L,x_R$ such that $\rho(x_{L,R})=0$. 
To summarise, we have an effective theory for the quantum hydrodynamical fluctuations  of the Fermi surface of  free fermions corresponding to the fluctuations of the eigenvalues of a random matrix $H$. They are described by  a 2D free boson BCFT on a curved geometry determined by the spectral density $\rho_0(E)$. In this formalism it is straightforward to consider double scaled matrix models, it is enough to use the double scaled spectral density which has only one zero at $x_L=0$ and $x_R=\infty$ so we have a BCFT on the half-line. \\

To complete the correspondence we are in need of a prescription to relate local operators $\mathcal{O}_{F}$ in the  microscopic fermionic theory (\ref{eq:fermionFT}), such as $\Psi,\Psi^{\dagger}$, to local operators  $O_{\text{Eff}}$ in the effective theory (\ref{eq:hydroaction}).
A single operator in the microscopic theory  $\mathcal{O}_F(x)$ can be expanded as a sum of local operators $O_{\text{Eff}}(x)$:
\begin{equation}
    \mathcal{O}_{F}(x)=\sum_i \tilde{A}_{ \mathcal{O},O_i} O_i(x),
\end{equation}
where $\tilde{A}_{ \mathcal{O},O_i}$ are dimensionful coefficients $[\tilde{A}_{ \mathcal{O},O_i}]=\Delta_{O_i}-\Delta_{\mathcal{O}}$ and we dropped the subscript from the  operators $O_{\text{Eff}}(x)$ in the right hand side.
We define dimensionless coefficients $A_{ \mathcal{O},O_i}$ using the characteristic length scale of the microscopic system  $d=(\rho_0(x))^{-1}$ , we have then 
\begin{equation}
    \tilde{A}_{ \mathcal{O},O_i}= \frac{A_{ \mathcal{O},O_i}}{\rho_0(x)^{\Delta_{O_i}-\Delta_{\mathcal{O}}}}.
\end{equation}
Using this prescription we can in principle write any correlation function in the microscopic model as a sum of CFT correlators which we know explicitly given the simplicity of the  CFT:
\begin{equation}
    \braket{ \mathcal{O}_F(x_1)\dots \mathcal{O}_F(x_n)}= \sum_{i_1,\dots, i_n} \frac{A_{ \mathcal{O},O_{i_1}}}{\rho_0(x_1)^{\Delta_{O_{i_1}}-\Delta_{\mathcal{O}}}} \dots \frac{A_{ \mathcal{O},O_{i_n}}}{\rho_0(x_n)^{\Delta_{O_{i_n}}-\Delta_{\mathcal{O}}}}\braket{O_{i_1}(x_1)\dots O_{i_n}(x_n)}_{CFT}.
\end{equation}
The sum can be organized according to the relevance of the  operators in the effective theory as each term in the sum is suppressed by the UV scale $d(x)$ as $ d^{\sum_k \Delta_{O_k}-n\Delta_{\mathcal{O}}}$.
The dimensionless coefficients $A_{ \mathcal{O},O_i}$ are   determined by matching to the microscopic theory, as usual in effective theories. \\
Consider the fermionic fields $\Psi^{\dagger},\Psi$ in the microscopic theory, they are charged under a global $U(1)$ symmetry with charge $q=1$ so the the corresponding CFT operators should also be.  The bosonic U(1) charge is the winding or magnetic number $q$ of vertex operators $V_{p,q}$.
Thus the corresponding operators are the CFT vertex operators  $V_{p,q=1}$ and their descendants, where we define a $(p,q)$ vertex operator by 
\begin{equation}
V_{p,q}(z,\bar{z})=:e^{i\qty(p-\frac{q}{2})\phi(z)+i\qty(p+\frac{q}{2})\bar{\phi}(\bar{z})}: .
\end{equation}
We have used chiral factorization of the CFT to write the boson field $h(x,t)$ as a sum of  holomorphic and antiholomorphic fields $h(x,\tau)=\phi(z)+\bar{\phi}(\bar{z})$.

Considering only the most relevant most relevant operator  we have then:
\begin{equation}
\label{Fermion-vertex}
\Psi^{\dagger}(x)\approx \tilde{A}_{\Psi^{\dagger},\mathcal{V}_{0,1}}(x)\mathcal{V}_{0,1}(x).
\end{equation}

Since the fermion fields have dimension $\frac{1}{2}$ and the  vertex operator has dimension $\frac{1}{4}$  the coefficient $\tilde{A}_{\Psi^{\dagger},\mathcal{V}_{0,1}}$ has dimensions  $-\frac{1}{4}$ such that:
\begin{equation}
\tilde{A}_{\Psi^{\dagger},\mathcal{V}_{0,1}}(x)=A_{\Psi^{\dagger},\mathcal{V}_{0,1}}\rho_0(x)^{1/4}.
\end{equation}
The dimensionless coefficient $A_{\Psi^{\dagger},\mathcal{V}_{0,1}}$ is the same as in the homogeneous $V(x)=0$ case, and so it can be calculated analytically by Bethe-Ansatz to obtain:

\begin{equation}
|A_{\Psi^{\dagger},\mathcal{V}_{0,1}}|^2=\frac{G^4(3/2)}{\sqrt{2\pi}},
\end{equation}
where $G$ indicates Barnes' $G$-function.
This completes the prescription for how to compute observables in the fermion theory using the hydrodynamical effective theory. 
We can now proceed to apply this framework to study the quantum mechanics of random matrix eigenvalues.

%% file: 3results.tex
\section{Spectral correlations and entanglement}\label{sec:CFT}
We now apply the effective theory describing the quantum hydrodynamical  fluctuations of the eigenvalue density  (eq. \ref{eq:hydroaction}) . The theory is a free boson 2D CFT on a non trivial background with boundaries. Thanks to its simplicity, we can easily compute many quantities of interest  straightforwardly and reproduce previous results obtained via less trivial methods. Let us first summarize the results we derive.\\
We compute corrections to the semiclassical spectral density $\rho_0(E)$ reproducing the leading non-perturbative correction to $\rho(E)$ as described, for example, in sec. 5 and appendix A of \cite{Saad:2019lba}. \\
We compute the two-level correlation between eigenvalues $\rho(E_1,E_2)$ and reproduce the ramp and plateau contributions to the Spectral Form Factor in the limit $|E_1-E_2|\ll1$. \\
We compute the spectrum of Renyi entropies  for an arbitrary interval bipartition of the eigenvalues $S_n(E_1,E_2)$ which generalises the results of \cite{Hartnoll:2015fca} to an arbitrary spectral curve $\rho_0(E)$ and reproduces their results in the appropriate limit. \\
Finally we compute the $n<N$ eigenvalue reduced density matrix  obtained by integrating out $(N-n)$ eigenvalues. \\
Let us start by defining a new coordinate $X$:
\begin{equation}
    X(x)=\int^x_{x_L}\frac{dx'}{\pi \rho_0(x')}, \qquad dX=\frac{dx}{\pi\rho_0(x)} 
\end{equation}
such that the metric $g_{ab}(x)$ becomes conformally flat:
\begin{equation}
    ds^2=\pi^2\rho_0(x)^2(dX^2+dt^2).
\end{equation}
The domain of the coordinate $X$ is $X\in (0,L)$, where $L\equiv X(x_R)$ is given by the mapping the right boundary point.
Since $\pi\rho_0(x)$ is the Fermi velocity, we can think of the coordinate $X(x)$ as the time it takes for an eigenvalue to go from the boundary $x_L$ to the point $x$.

\subsection{Spectral correlations}
\label{ssec:correlations}
We start by considering correlations of spectral densities $\braket{\rho(E)}$ and $\braket{\rho(E_1)\rho(E_2)}$.
 The spectral correlations in RMT are computed by averaging the discrete density $\rho(E)$ over the ensemble of random matrices $H$:
\begin{equation}
    \rho(E)\equiv \sum_{i=1}^N\delta(E-E_i) \rightarrow \braket{\rho(E)}_{H}=\int dH e^{\text{tr}V(H)}\rho(E).
\end{equation}
In Matrix Quantum Mechanics the average over random  matrices becomes a quantum expectation value of the density operator $\hat{\rho}(E)$: 
\begin{equation}
    \hat{\rho}(E)\equiv \Psi^{\dagger}(E)\Psi(E) \rightarrow \braket{\hat{\rho}(E)}.
\end{equation}

The $n$-eigenvalue correlation is then given by the $n$-point correlation function of the density operator. 
The quantum hydrodynamics effective theory allows us to easily compute these density correlations \cite{brundubailgff} in terms of free CFT correlation functions.  We are able to reproduce the leading non-perturbative corrections to $\braket{\rho(E)}$ and $\braket{\rho(E_1)\rho(E_2)}$ discussed in sec. 5 of \cite{Saad:2019lba} . This shows that we can think of matrix integrals as equal time expectation values in matrix quantum mechanics.
We start by expanding the  density operator  $\hat{\rho}(x)$ into CFT operators:
\begin{equation}
\hat{\rho}(x,t)\approx\rho_0(x) \text{Id}+\frac{\partial_xh(x,t)}{2\pi} + \sum^{\infty}_{p=1}\qty( A_{\rho,\mathcal{V}_{p,0}}\mathcal{V}_{p,0}(x,t)+A_{\rho,\mathcal{V}_{-p,0}}\mathcal{V}_{-p,0}(x,t)).
\end{equation}

The first two operators follow directly from the construction of the effective theory since $ \rho_0(x)$ is the  saddle point value of the density and  the linear variation around the saddle is $\delta\rho\equiv \frac{\partial_xh(x,t)}{2\pi}$. The expansion of $\hat{\rho}$ only includes vertex operators $\mathcal{V}_{p,q}$ with $q=0$ since $\hat{\rho}$ does not change the total number of eigenvalues. 
Keeping only the most relevant operators in the expansion we have:
\begin{equation}
\hat{\rho}(x,t)\approx\rho_0(x) \text{Id}+\frac{\partial_xh(x,t)}{2\pi} + A_{\rho,\mathcal{V}_{1,0}}\mathcal{V}_{1,0}(x,t)+A_{\rho,\mathcal{V}_{-1,0}}\mathcal{V}_{-1,0}(x,t).
\end{equation}
The coefficients $A_{\rho,\mathcal{V}_{\pm1,0}}$ are naturally dimensionless since both  $\hat{\rho}$ and $\mathcal{V}_{\pm 1,0}$ have scaling dimension $\Delta=1$. 

They are given by the following expression
\begin{equation}
A_{\rho,\mathcal{V}_{\pm1,0}} =\frac{1}{2\pi} e^{ \pm 2 \pi i \theta(x)}, \qquad \theta(x)=\int_0^x \rho_0(x')dx' -\frac{1}{2}.
\end{equation}
The absolute value $|A_{\rho,\mathcal{V}_{\pm1,0}}|=\frac{1}{2\pi}$  can be obtained exactly from Bethe-Ansatz form factors (see appendix B of \cite{brundubailgff}).  The phase $\theta(x)$ is a WKB phase. 

\subsubsection{Non perturbative corrections to density of eigenvalues $\braket{\rho(E)}$}
The quantum hydrodynamical fluctuations will give the leading non-perturbative corrections to the semiclassical spectral density $\rho_0(x)$. 
These corrections produce oscillations on top of the semiclassical density which, from the fermionic point of view, can be identified as Friedel oscillations.
Let us now compute $\braket{\hat{\rho}(x)}$ by taking  the expectation value of the previous expression for $\hat{\rho}$. We have that $\braket{\partial_x h}=0$ due to $\mathbb{Z}_2$ symmetry. The vertex operator VEV is obtained again by first changing coordinates $X(x)$ so the metric is conformally flat, performing a Weyl transformation to go to flat space and using a conformal transformation $w(z)=e^{i\frac{\pi}{L}z}$ to map the strip to the upper half plane $\mathbb{H}$:
\begin{equation}
\braket{\mathcal{V}_{\pm 1,0}(z)}_{g}=(\pi\rho_0(x))^{-1}\braket{\mathcal{V}_{\pm 1,0}(z)}_{strip}=(\pi\rho_0(x))^{-1}\qty|\frac{dw}{dz}|\braket{\mathcal{V}_{\pm 1,0}(w(z))}_{\mathbb{H}} .
\end{equation}
The expectation value on the upper half plane can be computed by the method of images and is given by 
\begin{equation}
    \braket{\mathcal{V}_{\pm 1,0}(w(z))}_{\mathbb{H}}= e^{\frac{1}{2}G_D(w)} \qquad G_D(w)= -\log|w-\bar{w}|^2,
\end{equation}

 where $G_D(w)$ is the regularised Green's function at coincident points with Dirichlet boundary conditions. Setting $t=0$ in $z=X+it$ and using the mapping $w(z)=e^{i\frac{\pi}{L}z}$ we have: 
 \begin{equation}
     G_D(X)=-\log\qty|2\sin(\frac{\pi}{L}X)|^2.
 \end{equation}
We arrive at the following expression for $\braket{\hat{\rho}}$:

\begin{equation}\label{eq:densityosc}
\braket{\hat{\rho}(x)}=\rho_0(x)-\frac{\cos(2\pi \int^x\rho_0(x')dx')}{2\pi L \rho_0(x) \sin(\frac{\pi X}{L}) }.
\end{equation}
As a consistency check, this expression precisely matches with the large $N$ limit of the exact solution of the Gaussian matrix model, given by Hermite polynomials.  \footnote{For the GUE the map to free fermions in a harmonic potential is actually exact \cite{Dean_2019}, since  the probability density $|\psi_N(x_1,\dots, x_N)|^2$ is equal to the GUE joint eigenvalue probability density $|\psi_N(x_1,\dots, x_N)|^2=\rho_{GUE}(x_1,\dots,x_N)$. }
As a function of the eigenvalues $E$ the density is:
\begin{equation}
\braket{\hat{\rho}(E)}=\rho_0(E)-\frac{\cos(2\pi \int^E\rho_0(E')dE')}{2\pi L \rho_0(E) \sin(\frac{1}{L} \int^E \frac{dE'}{\rho_0(E')}) },
\end{equation}
This gives the first quantum correction to the spectral density $\rho(E)$.
In a double scaled matrix model, where $L=\infty$, we have:
\begin{equation}
\braket{\hat{\rho}(E)}=\rho_0(E)-\frac{\cos(2\pi \int^E\rho_0(E')dE')}{2\pi  \rho_0(E) \qty(\int^E \frac{dE'}{\rho_0(E')}) },
\end{equation}
This is a non-perturbative correction to the density of states since it is of the form $\cos(e^{S_0})=e^{ie^{S_0}}$. It reproduces the leading non-perturbative correction to the density of states.\\
As an example, take the Airy spectral curve $\rho_0(E)=\sqrt{E}$ for which we obtain: 
 \begin{equation}
\braket{\hat{\rho}(E)}=\rho_0(E)-\frac{\cos(2\pi \int^E\rho_0(E')dE')}{4\pi E},
\end{equation}
this is exactly the expression in eq. (155) of \cite{Saad:2019lba}.

\subsubsection{The ramp and plateau in $\braket{\rho(E_1)\rho(E_2)}$}
We can now compute the two-point function of the density of eigenvalues $\braket{\hat{\rho}(E_1)\hat{\rho}(E_2)}$. One has to multiply the expansions for the density operators and take the expectation value. Two point functions of vertex operators and the height field are entirely determined in terms of the  Green function with Dirichlet boundary conditions $G_D(w_1,w_2)$ on $\mathbb{H}$ which is given by: 
\begin{equation}
    G_D(w_1,w_2)=\log\qty|\frac{w_1-w_2}{w_1-\bar{w}_2}|^2 .
\end{equation}
Evaluating it at equal times $t_1=t_2=0$ and using the mapping $w(z)=e^{i\frac{\pi}{L}z}$ we have: 
\begin{equation}
    G_D(\bar{X}_1,\bar{X}_2)=\log\qty|\frac{\sin(\frac{\bar{X}_1-\bar{X}_2}{2})}{\sin(\frac{\bar{X}_1+\bar{X}_2}{2})}|^2, \qquad \bar{X}=\frac{\pi}{L}X.
\end{equation}

The two-point correlation of the spectral density is given by:

\begin{equation}
    \begin{split}
       \braket{\hat{\rho}(x_1) \hat{\rho}(x_2)}_{\mathrm{c}}&=\frac{1}{\pi^2\rho_0(x_1)\rho_0(x_2)} \Bigg\{ -\frac{\partial_{\bar{X}_1} \partial_{\bar{X}_2} G_{D}(\bar{X}_1, \bar{X}_2) }{4\pi^2}
+\\&\qty[\partial_{\bar{X}_1} G_D(\bar{X}_1, \bar{X}_2) \sin(2 \pi\theta(x_2)) e^{\frac{1}{2} G_D(X_2)}+(\bar{X}_1 \leftrightarrow \bar{X}_2)]+
\\&
e^{\frac{1}{2}(G_D(X_1)+G_D(X_2))}\left(e^{G_D(\bar{X}_1, \bar{X}_2)}-1\right) \cos \left[2 \pi(\theta(x_1)+\theta(x_2))\right]+\\&e^{\frac{1}{2}(G_D(X_1)+G_D(X_2))}\left(e^{-G_D(\bar{X}_1, \bar{X}_2)}-1\right) \cos \left[2 \pi(\theta(x_1)-\theta(x_2))\right]  \Bigg\}.
    \end{split}
\end{equation}

The derivatives appearing in the expression are the following:
\begin{equation}
\partial_{\bar{X}_1}G_D(\bar{X}_1,\bar{X}_2)=\qty[\cot(\frac{\pi(X_1-X_2)}{2L}) -\cot(\frac{\pi(X_1+X_2)}{2L})]
\end{equation}
\begin{equation}
\partial_{\bar{X}_2}G_D(\bar{X}_1,\bar{X}_2)=-\qty[\cot(\frac{\pi(X_1-X_2)}{2L}) +\cot(\frac{\pi(X_1+X_2)}{2L})]
\end{equation}
\begin{equation}
\partial_{\bar{X}_1}\partial_{\bar{X}_2}G^D(\bar{X}_1,\bar{X}_2)=\frac{1}{2}\qty[\sin^{-2}\qty(\frac{\pi(X_1-X_2)}{2L}) +\sin^{-2}\qty(\frac{\pi(X_1+X_2)}{2L})].
\end{equation}
For finite $L$ it is enough to substitute the expressions for the Green functions and their derivatives, which we won't do explicitly.  For a double scaled matrix model where $L\rightarrow \infty$ we have:

\begin{equation}
\begin{split}
    \braket{\hat{\rho}(x_1)\hat{\rho}(x_2)}=& \frac{1}{2\pi^2}\qty(\frac{1}{\pi^2\rho_0(x_1)\rho_0(x_2)})\Bigg\{ -\qty(\frac{1}{(X_1-X_2)^2}+\frac{1}{(X_1+X_2)^2}) + \\& \qty[\frac{\sin(2\pi \theta(x_2))}{X_2}\qty(\frac{1}{(X_1-X_2)}-\frac{1}{(X_1+X_2)})+(X_1 \leftrightarrow X_2)]+\\& \qty(-\frac{\cos(2\pi (\theta(x_1)+\theta(x_2)))}{(X_1+X_2)^2}+\frac{\cos(2\pi (\theta(x_1)-\theta(x_2)))}{(X_1-X_2)^2})\Bigg\}.
    \end{split}
\end{equation}
Approximating the integral $\int_0^E\frac{1}{\rho_0(E')}\approx \frac{E}{\rho_0(E)}$ we have that $X\approx \frac{E}{\pi \rho_0(E)}$. Moreover, in the limit where $|E_2-E_1|\ll 1$ we can write $\rho_0(E_1)=\rho_0(E_2)=\rho_0(E)$ where $E=\frac{E_1+E_2}{2}$. 
We have then:
\begin{equation}
\begin{split}
    \braket{\hat{\rho}(E_1)\hat{\rho}(E_2)}=& \frac{1}{2\pi^2}\Bigg\{ -\qty(\frac{1}{(E_1-E_2)^2}+\frac{1}{(E_1+E_2)^2}) + \\&  
    \frac{1}{(E_1-E_2)}\qty(\frac{\cos(2\pi\int^{E_2}\rho_0(E')dE')}{E_2}-\frac{\cos(2\pi\int^{E_1}\rho_0(E')dE')}{E_1})+\\& \qty(-\frac{\cos(2\pi \qty(\int^{E_2}\rho_0(E')dE'+\int^{E_1}\rho_0(E')dE'))}{(E_1+E_2)^2}+\frac{\cos(2\pi \int_{E_1}^{E_2}\rho_0(E')dE')}{(E_1-E_2)^2})\Bigg\}.
    \end{split}
\end{equation}
For $|E_2-E_2|\ll1$ we recover the well known Sine kernel expression for the two-point correlation of eigenvalues in Random Matrix Theory: 
\begin{equation}
\begin{split}
    \braket{\hat{\rho}(E_1)\hat{\rho}(E_2)}&= -\frac{1}{2\pi^2} \frac{1}{(E_1-E_2)^2} +\frac{1}{2\pi^2}\frac{\cos(2\pi \int_{E_1}^{E_2}\rho_0(E')dE')}{(E_1-E_2)^2} + \text{reg}\\&= -\frac{1}{\pi^2}\frac{\sin^2(\pi \int_{E_1}^{E_2}\rho_0(E')dE')}{(E_1-E_2)^2}.
    \end{split}
\end{equation}
We have reproduced the known matrix integral expressions for $\braket{\rho(E)}$ and $\braket{\rho(E_1)\rho(E_2)}$ by considering equal time quantum expectation values of the eigenvalue density operator $\hat{\rho}(E)$ in Matrix Quantum Mechanics. This shows that, in this sense, we can think of a matrix integral as a fixed time instance of a corresponding quantum mechanical theory of matrices.
The statistical fluctuations given by integrating over an ensemble of matrices can now be understood as quantum fluctuations of a single matrix in the large $N$ limit.

\subsection{Entanglement entropy}\label{ssec:entent}
An observable present in MQM that has no analogue in RMT is the entanglement between eigenvalues. Since the eigenvalues are quantum mechanical with a collective wavefunction $\Psi_N(E_1,\dots, E_N)$ we can consider the entanglement entropy for a bipartition of eigenvalue space. 
Thanks to the CFT description of the hydrodynamical fluctuations we  can compute the entanglement entropy using the universal Cardy-Calabrese formula \cite{cardycalab}. Thus we  don't need to map any microscopic operators to the effective theory in this case. 
The Renyi entropies for a subsystem $A$ are defined as:
\begin{equation}
    S_n\equiv\frac{1}{1-n}\log(\text{Tr}(\rho_A^n)),
\end{equation}
where $\rho_A$ is the reduced density matrix.
\subsubsection*{Half-space bipartition $(0,E)\cup (E,\infty)$}
We consider the Renyi entropies for a bipartition $(x_L,x) \cup (x,x_R)$ which in $X$ coordinates is $(0,X(x))\cup(X(x),L)$.
In a 2D CFT the Renyi entropies for such a bipartition can be computed by the expectation value of a single twist field \cite{cardycalab}:
\begin{equation}
S_n(x)=\frac{1}{1-n}\log(\epsilon^{\Delta_n}\braket{\mathcal{T}_n(x,t=0)}),
\end{equation}
where $\Delta_n$ is the dimension of the twist operator:
\begin{equation}
\Delta_n=\frac{c}{12}(n-\frac{1}{n}),
\end{equation}
and $\epsilon$ is a UV cut-off for the formally divergent entanglement entropy. The cut-off is known to encode the divergent amount of short-range entanglement in continuum Quantum Field Theories. This divergence is an issue when attempting to give a rigorous definition of entanglement entropy in QFT (see \cite{witten1, witten2}). However, as illustrated in section 2, in the  effective hydrodynamical description we have a microscopic length scale, the mean particle spacing $d(x)=\rho_0^{-1}(x)$ which gives a natural UV cut-off $\Lambda_{UV}=\rho_0(x)$. Since the system is inhomogeneous, the UV cutoff scale is position dependent and we have:
\begin{equation}
    \epsilon(x)=\frac{\epsilon_0}{\rho_0(x)},
\end{equation}
 where $\epsilon_0$ is a dimensionless constant and the cutoff is evaluated at  the boundary point $x$ of the bipartition. We have then a UV-finite expression for the entanglement entropy. This can be interpreted as a consequence of having a finite density of eigenvalues $\rho_0(E)$ in RMT due to eigenvalue repulsion. Equivalently, it is a consequence of having a non-zero VEV for the  density $\rho(x)=\braket{\Psi^{\dagger}\Psi}$. In the next section we will see that in the duality between the $c=1$ matrix model and two-dimensional string theory the finiteness of the entanglement entropy can be interpreted as due to stringy effects \cite{Hartnoll:2015fca}. String theory, as expected, regulates the UV-divergence to give a finite answer for the entropy.

We work in complex coordinates $z=X+it$ defined on the infinite strip $(0,L) \times \mathbb{R}$. The metric in complex coordinates is 
$ds^2=\pi^2\rho_0(x)^2 dzd\bar{z}$. We can perform a Weyl transformation $g_{ab}\rightarrow (\pi\rho_0(x))^{-2}g_{ab}$ to go to flat space.  Under the Weyl transformation the twist operator scales as  $\mathcal{T}_n\rightarrow (\pi\rho_0(x))^{-\Delta_n}\mathcal{T}_n$. 
Next we  map the $z$-strip to the upper half-plane $\mathbb{H}$ via a conformal transormation $g(z)=e^{i\pi\frac{z}{L}}$. Under this map  the twist field transforms as $\mathcal{T}_n(z)_{\text{Strip}}\rightarrow \qty|\dv{g(z)}{z}|^{\Delta_n} \mathcal{T}_n(g(z))_{\mathbb{H}}$.  The last ingredient is the expectation value of the twist field on the upper-half plane which is $\braket{\mathcal{T}_n(g(z))}_{\mathbb{H}}=(\text{Im}g(z))^{-\Delta_n}$.
We combine everything together to arrive at:
\begin{equation}
S_n(x)=\frac{n+1}{12n}\log \qty[\frac{\Omega(x)}{\epsilon(x)}\qty|\dv{g(z)}{z}|^{-1}\text{Im}g(z)],
\end{equation}
where
\begin{equation}
\qty|\dv{g(z)}{z}|= \frac{\pi}{L},
\end{equation}
\begin{equation}
\text{Im}g(z)=\sin(\frac{\pi X}{L}).
\end{equation}
We then obtain the entanglement entropy for a bipartition $(x_L,x)\cup(x,x_R)$ of  eigenvalues in a model with spectral density $\rho_0(x)$:
\begin{equation}
S_n=\frac{n+1}{12n}\log(\frac{L}{\pi} \rho_0^2(x)\sin(\frac{\pi X(x)}{L}))+\text{const}.
\end{equation}
Writing this explicitly in the eigenvalue coordinate $x=E$ for a bipartition $(E_L,E)\cup(E,E_R)$ we have: 

\begin{equation}
S_n(E)=\frac{n+1}{12n}\log(\frac{L}{\pi} \rho_0^2(E)\sin(\frac{1 }{L} \int_{E_L}^E \frac{dE'}{\rho_0(E')} ))+\text{const}.
\end{equation}

We see that, due to the term $\sin(\frac{\pi}L X)$ with $X\in(0,L)$, the entropy has the expected Page-curve behaviour as a function of the bipartition point $E$, increasing until a turning point and decreasing afterwards. This is a consequence of unitarity in a system with a finite number of states $N=\int \rho_0(E)dE$ and does not survive the double scaling limit. As $N\rightarrow \infty$ in the double scaling limit we lose unitarity of the entanglement entropy $S_n(E)$ and we have information loss. There are many examples of the tension between unitarity  and the large $N$ limit \cite{Maldacena:2001kr,witten1,Schlenker:2022dyo,Witten:2021unn,Leutheusser:2021frk, Chandrasekaran:2022eqq} . 

For a double scaled matrix model we have that $x_R=\infty$ and thus $L=\infty$. Expanding the sine we have that the $L$ dependence drops out and we obtain:
\begin{equation}
S_n(E)=\frac{n+1}{12n}\log(\frac{1}{\pi} \rho_0^2(E) \int_{E_L}^E \frac{dE'}{\rho_0(E')} )+\text{const}.
\end{equation}
Approximating the integral by assuming an almost constant density  $\int_{E_L}^E \frac{dE'}{\rho_0(E')}\approx \frac{E-E_L}{\rho_0(E)}$ we obtain: 
\begin{equation}
S_n(E)\approx\frac{n+1}{12n}\log(\frac{1}{\pi} \rho_0(E) (E-E_L) )+\text{const}.
\end{equation}
We see that the entanglement entropy  presents a leading term   proportional to the microcanonical entropy: 
\begin{equation}
    S_n(E) \propto \log(\rho_0(E))=S_{0}(E).
\end{equation}

This shows that the entanglement entropy contains a large amount of short-range entanglement coming from neighbouring eigenvalues separated by the bipartition, since the  entropy $S_0(E)$ counts the number of states in a window $(E-dE,E+dE)$.\\ 
The second term $S_n(E)\propto \log(E-E_L)$ is the usual Cardy-Calabrese behaviour in two dimensions. Note that we should not extrapolate this result to the edge of the spectral density $E\sim E_L$. In RMT, there are distinct 'bulk' and 'edge' limits for the spectral density with different universal kernels describing them \cite{Tao} and we should not extrapolate a bulk result to the edge region. Hydrodynamics requires  the derivative of the density to be small, thus it describes the 'bulk' region.  
\
\subsubsection*{Interval bipartition $(E_1,E_2)$}
We now calculate the  Renyi entropies for a subregion $A=(X_1,X_2)\in (0,L)$ equal to $A=(x_1,x_2)\in (x_L,x_R)$, where $X_{1,2}\equiv X(x_{1,2})$. The Renyi entropies are given by the two point function of the twist field at the extrema of the interval:
\begin{equation}
S_n\propto\log(\epsilon^{\Delta_n}\braket{\mathcal{T}_n(z_1)\tilde{\mathcal{T}}_n(z_2)}).
\end{equation}
$\tilde{\mathcal{T}}_n$ is the conjugate twist field \cite{cardycalab}.
The cutoff now is  $\epsilon=\frac{\epsilon_0}{\rho(x_1)\rho(x_2)}$. Repeating the same manipulations as before we can reduce the computation to a correlation on the flat upper half plane $\mathbb{H}$:
\begin{equation}
\braket{\mathcal{T}_n(z_1)\tilde{\mathcal{T}}_n(z_2)}_{\text{strip}}=\qty(\Omega(x_1)\qty|\dv{g(z_1)}{z_1}|)^{\Delta_n}\qty(\Omega(x_2)\qty|\dv{g(z_2)}{z_2}|)^{\Delta_n} \braket{\mathcal{T}_n(g(z_1))\tilde{\mathcal{T}}_n(g(z_2))}_{\mathbb{H}}.
\end{equation}

The two point function on the upper half plane $\mathbb{H}$ is a priori a non trivial calculation corresponding to a four-point function on the plane. In the case of a free boson theory it is known \cite{casini,cardycalab, DiFrancesco:1997nk} and we have:
\begin{equation}
\braket{\mathcal{T}_n(z_1)\tilde{\mathcal{T}}_n(z_2)}=\qty[\Omega(x_1)\qty|\dv{g(z_1)}{z_1}|\Omega(x_2)\qty|\dv{g(z_2)}{z_2}| \frac{|g^*(z_1)-g(z_2)|^2}{\text{Im}g(z_1)\text{Im}g(z_2)|g(z_1)-g(z_2)|^2}]^{\Delta_n}.
\end{equation}
The  Renyi entropies are: 
\begin{equation}
S_n\propto \log\qty[\qty(\frac{L}{\pi})^2\rho_0^2(x_1)\rho_0^2(x_2)\sin(\pi \frac{X_1}{L} ) \sin(\pi \frac{X_2}{L}) \frac{|g(z_1)-g(z_2)|^2}{|g^*(z_1)-g(z_2)|^2}],
\end{equation}
where the last term is equal to:
\begin{equation}
\frac{|g(z_1)-g(z_2)|^2}{|g^*(z_1)-g(z_2)|^2}=\frac{1-\cos(\frac{\pi}{L}(X_2-X_1))}{1-\cos(\frac{\pi}{L}(X_1+X_2))}=\frac{\sin^2(\frac{\pi}{2L}(X_2-X_1))}{\sin^2(\frac{\pi}{2L}(X_1+X_2))}.
\end{equation}
The final result is then:
\begin{equation}
   S_n(x_1,x_2)=\frac{n+1}{12n} \log(\qty(\frac{L}{\pi})^2\rho_0^2(x_1)\rho_0^2(x_2)\sin(\pi \frac{X_1}{L} ) \sin(\pi \frac{X_2}{L})\frac{\sin^2(\frac{\pi}{2L}(X_2-X_1))}{\sin^2(\frac{\pi}{2L}(X_1+X_2))})+\text{const}.
\end{equation}
Writing this in terms of the eigenvalues $(E_1,E_2)$ we have:
\begin{equation}
\begin{split}
   S_n(x_1,x_2)=\frac{n+1}{12n} \log\Bigg[\qty(\frac{L}{\pi})^2&\rho_0^2(E_1)\rho_0^2(E_2)\sin( \frac{1}{L} \int_{E_L}^{E_1} \frac{dE'}{\rho_0(E')} ) \sin( \frac{1}{L}\int_{E_L}^{E_2} \frac{dE'}{\rho_0(E')} )\\&\times\frac{\sin^2\qty(\frac{1}{2L}\int_{E_1}^{E_2} \frac{dE'}{\rho_0(E')})}{\sin^2\qty(\frac{1}{2L}\qty(2\int_{E_L}^{E_1} \frac{dE'}{\rho_0(E')}+\int_{E_1}^{E_2} \frac{dE'}{\rho_0(E')}))}\Bigg]+\text{const}.
   \end{split}
\end{equation}
For a double scaled matrix model $L=\infty$ we have: 
\begin{equation}
   S_n(x_1,x_2)=\frac{n+1}{12n} \log(\rho_0^2(x_1)\rho_0^2(x_2)X_1X_2 \frac{(X_2-X_1)^2}{(X_1+X_2)^2})+\text{const},
\end{equation}
which in terms of the eigenvalues is:
\begin{equation}
\begin{split}
   S_n(E_1,E_2)=\frac{n+1}{12n} \log\Bigg[\qty(\frac{1}{\pi})^2&\rho_0^2(E_1)\rho_0^2(E_2) \qty(\int_{E_L}^{E_1} \frac{dE'}{\rho_0(E')} ) \qty( \int_{E_L}^{E_2} \frac{dE'}{\rho_0(E')} )\\&\times\frac{\qty(\int_{E_1}^{E_2} \frac{dE'}{\rho_0(E')})^2}{\qty(2\int_{E_L}^{E_1} \frac{dE'}{\rho_0(E')}+\int_{E_1}^{E_2} \frac{dE'}{\rho_0(E')})^2}\Bigg]+\text{const}.
   \end{split}
\end{equation}
Approximating the integrals by assuming an almost constant density  $\int_{E_L}^E \frac{dE'}{\rho_0(E')}\approx \frac{E-E_L}{\rho_0(E)}$ we obtain:
\begin{equation}
   S_n(E_1,E_2)\approx\frac{n+1}{12n} \log\Bigg[\qty(\frac{1}{\pi})^2\rho_0(E_1)\rho_0(E_2)(E_1-E_L)(E_2-E_L)\frac{(E_2-E_1)^2}{\qty((E_2+E_1)-2E_L)^2}\Bigg] 
\end{equation}
 In the limit of RMT unviersality where $(E_1-E_2)\ll1$ we obtain the simple expression: 
 \begin{equation}
   S_n(E_1,E_2)\approx\frac{n+1}{12n} \log\qty(\frac{\rho_0^2(E)(E_1-E_2)^2}{4\pi^2})
\end{equation}
where $E=\frac{E_1+E_2}{2}$ is the average energy. We see again that there is a leading contribution proportional to the average  microcanonical entropy $S_n\propto S_0(E)=\log(\rho_0(E))$ of the interval $(E_1,E_2)$. 
The entanglement entropy derived via the hydrodynamic CFT has been checked against numerical simulations in \cite{dubailcalab} for a double well potential finding excellent agreement. 
\subsubsection{Emergence of spacetime in 2D string theory}

Two-dimensional string theory is a non-critical bosonic string theory in $D=2$ flat spacetime with a linear dilaton background and a massless tachyon. The worldsheet is a Liouville CFT with $c_L=25$ and a free boson $c_M=1$ which cancel the ghost central charge $c_g=-26$ \cite{martinec,klebanov}. The low energy effective action is \cite{Callan:1985ia,Natsuume:1994sp}:
\begin{equation}
    S=\frac{1}{2}\int dt dx \sqrt{-g}e^{-2\Phi}\qty(\frac{R}{2}+2(\partial\Phi)^2 +8 - (\partial T)^2+4T^2-2V(T)),
\end{equation}
where $V(T)$ is a potential for the tachyon. The background has a tachyon condensate with a free parameter $\bar{\mu}$ which determines the effective string coupling $g_{eff}\sim \bar{\mu}^{-1}$, thus we have a perturbative string theory for $\mu\gg1$. Equivalently, $\bar{\mu}$ is the cosmological constant of the worldsheet Liouville theory.
Two-dimensional string theory is dual to a theory of matrix quantum mechanics with an inverted oscillator potential $V(x)=-\frac{x^2}{2}$ and chemical potential $\mu=-\bar{\mu}<0$.   The potential arises from  a double scaling limit of a potential  $U(x)=-\frac{x^2}{2}+gx^3$ by taking $N\rightarrow \infty$ and $g\rightarrow 0 $ while keeping $\mu=-N\epsilon_F$ fixed, where $\epsilon_F$ is the Fermi energy of the fermions filling up the well in the potential. This zooms into the vicinity of the maximum of the potential which gives the universal critical behaviour. The supersymmetric string is described instead by a double well potential  \cite{Douglas:2003up}.  In the hydrodynamic approach it is perfectly possible to treat both the general case with potential $U(X)$ but we will focus on the inverted oscillator. The leading density of eigenvalues is given by \ref{eq:lda}: 
\begin{equation}
    \rho_0(x)=\frac{1}{\pi}\sqrt{2(-V(x)-\bar{\mu})}=\frac{1}{\pi}\sqrt{x^2-2\bar{\mu}}.
\end{equation}
It has a left edge $x_L=\sqrt{2\bar{\mu}}$ and extends to infinity as it is a double scaled model. 
The density of states gives the geometry of the Fermi  surface on which the 2D CFT ($\ref{eq:hydroaction}$) describing the quantum hydrodynamical fluctuations lives: 
\begin{equation}
    ds^2= (x^2-2\bar{\mu})d\tau^2 +dx^2.
\end{equation}
The coordinate transformation which renders the metric conformally flat is explicitly: 

\begin{equation}\label{eq:tof}
    X(x)=\int_{\sqrt{2\bar{\mu}}}^x \frac{d\lambda}{\sqrt{-2\bar{\mu}+\lambda^2}}=\cosh^{-1}\qty(\frac{x}{\sqrt{2\bar{\mu}}})= \log(\frac{x+\sqrt{x^2-2\bar{\mu}}}{\sqrt{2\bar{\mu}}}),
\end{equation}

It was argued first in \cite{Das:1995vj} that the transformation $X(x)$ gives, at the semiclassical level, the map from the eigenvalue-space $x$  to the string theory spacetime $X$ since physically it is the "time-of-flight", meaning the WKB time it takes for an eigenvalue to go from $x_L$ to a point $x$. Then in \cite{Hartnoll:2015fca} a consistency argument identifying $X(x)$ with the map between the eigenvalues and the emergent spacetime  was given based on entanglement. They computed the entanglement entropy of the eigenvalues using the techniques of \cite{Calabrese:2011ycz} and the WKB wavefunctions. Then they argued that the entanglement entropy in spacetime should be given by the Cardy-Calabrese formula and found that the relation $X(x)$ produced the right matching. 
The hydrodynamic approach instead provides a natural and more constructive point of view: the geometry of the Fermi surface is dual to the spacetime geometry and the mapping $X(x)$ is simply the map between the two metrics. The Cardy-Calabrese formula follows immediately from the fact that the  quantum hydrodynamical theory of the eigenvalues is a 2D CFT. We will now show explicitly the match with the entanglement entropy derived in \cite{Hartnoll:2015fca}. 
We can invert eq. \ref{eq:tof} to write the density $\rho_0(X)$ in spacetime coordinates: 
\begin{equation}
\pi^2\rho_0^2(x)=-2\mu +x^2=2\mu(-1+\cosh^2(X))=2\mu \sinh^2(X).
\end{equation}
One can define a string coupling given by: 
\begin{equation}
\frac{1}{\tilde{g}_s(X)}\equiv 2\bar{\mu} \sinh^2(X)\equiv \pi^2\rho_0^2(X),
\end{equation}
which at weak coupling $\bar{\mu}\gg1$ is equal to the string coupling in the linear dilaton background $\tilde{g}_s(X)=\frac{g_s(X)}{2\mu}$. 
The entanglement entropy for a spacetime bipartition $(0,X)U(X,L)$  in two-dimensional string theory is then: 
\begin{equation}
S_n(X)\Big|_{\text{2D  String }}=\frac{n+1}{12n}\log\qty[\frac{L}{\pi}\frac{1}{\tilde{g}_s(X)}\sin(\frac{\pi X}{L})] +\text{const},
\end{equation}
while for a spacetime interval $(X_1,X_2)$ it is given by:

\begin{equation}
S_n= \frac{n+1}{12n} \log(\qty(\frac{L}{\pi})^2 \frac{\sin(\pi \frac{X_1}{L} ) \sin(\pi \frac{X_2}{L})}{\tilde{g}_s(X_1)\tilde{g}_s(X_2)} \frac{\sin^2(\frac{\pi}{2L}(X_2-X_1))}{\sin^2(\frac{\pi}{2L}(X_1+X_2))}) +\text{const}.
\end{equation}
This reproduces the results of   \cite{Hartnoll:2015fca} when considering the Von Neumann entropy $n=1$ and  $L\rightarrow \infty$:
\begin{equation}
S_1= \frac{1}{6} \log\qty[ \frac{X_1 X_2}{\tilde{g}_s(X_1)\tilde{g}_s(X_2)} \frac{(X_2-X_1)^2}{(X_1+X_2)^2}]+\text{const}=\frac{1}{3}\log(\frac{X_2-X_1}{\sqrt{\tilde{g}_s(X_1)\tilde{g}_s(X_2)}}) +\frac{1}{6}\log(\frac{X_1 X_2}{X_1+X_2)^2}) +\text{const}.
\end{equation}
Since they were working  in the microscopic fermion field theory (\ref{eq:fermionFT}), they were able to determine the additive constant for $L\rightarrow \infty$. However the constant can depend on the system size $L$ so we cannot use their results to fix the constant in the general $L$ case.
Often in  in two-dimensional string theory a cut-off is introduced for the inverted oscillator potential which comes from the potential $U(x)$ before the double-scaling limit. The cut-off is at a distance $x_R\sim \frac{1}{g}\sim N$ so at large $N$ it is effectively not there.  
Since our results are valid for finite $L$ we can also probe the region close to the cut-off or directly do the computation for the potential $U(x)$.

\subsection{Reduced density matrix for $n<N$ eigenvalues}\label{ssec:dm}
The one-particle density matrix in the fermionic field theory \ref{eq:fermionFT} is computed by the two-point function of the fermion field \cite{abrikosov2012methods,PhysRevLett.47.1840,brundubaildm}:
\begin{equation}
    g_1(x,x')\equiv\braket{\Psi^{\dagger}(x)\Psi(x')}.
\end{equation}
We have seen that the fermionic operators $\Psi,\Psi^{\dagger}$ can be expanded as an infinite sum of CFT primary operators and their descendants consistent with the symmetries.  In particular, $\Psi,\Psi^{\dagger}$ corresponds to vertex operators $\mathcal{V}_{p,\pm1}$ and their derivatives.
Considering only the most relevant most relevant operator  we have:
\begin{equation}
\label{Fermion-vertex}
\Psi^{\dagger}(x)\approx A_{\Psi^{\dagger},\mathcal{V}_{0,1}}\rho_0(x)^{1/4} \mathcal{V}_{0,1}(x),
\end{equation}
where the dimensionless coefficient is given by 
\begin{equation}
|A_{\Psi^{\dagger},\mathcal{V}_{0,1}}|^2=\frac{G^4(3/2)}{\sqrt{2\pi}}.
\end{equation}

The one-eigenvalue reduced  density matrix is then given, at leading order in the hydrodynamic effective theory  simply by a two-point function of vertex operators:
\begin{equation}
    g_1(x,x')=|A_{\psi,\mathcal{V}_{0,-1}}|^2 \rho_0(x)^{1/4}\rho_0(x')^{1/4} \braket{\mathcal{V}_{0,1}(x),\mathcal{V}_{0,-1}(x')}_{\text{CFT}}.
\end{equation}
 We make use of coordinates  $X(x)$ such that the geometry is  conformally flat.
We map the correlator to the  flat space infinite strip:
\begin{equation}
\braket{\mathcal{V}_{0,1}(x),\mathcal{V}_{0,1}(x')}_{g}=\qty(\frac{dX}{dx})^{1/4} \qty(\dv{X}{x})^{1/4} \braket{\mathcal{V}_{0,1}(X),\mathcal{V}_{0,1}(X')}_{ flat}.
\end{equation}
The two-point function of vertex operators on an infinite  strip $(0,L)\times \mathbb{R}$ is known to be:
\begin{equation}
\braket{\mathcal{V}_{0,1}(X),\mathcal{V}_{0,1}(X')}_{CFT, flat}= \frac{\qty|\sin(\frac{\pi X}{L})\sin(\frac{\pi X'}{L})|^{1/4}}{\qty|\frac{2L}{\pi}\sin(\frac{\pi (X-X')}{2L})\sin(\frac{\pi (X+X')}{2L})|^{1/2}}.
\end{equation}
 We arrive at the following result for the one-eigenvalue density matrix: 
\begin{equation}
g_1(x,x')= \frac{|A_{\psi,\mathcal{V}_{0,-1}}|^2}{\sqrt{\pi}} \frac{\qty(\sin(\frac{\pi X}{L})\sin(\frac{\pi X'}{L}))^{1/4}}{\qty|\frac{2L}{\pi}\sin(\frac{\pi (X-X')}{2L})\sin(\frac{\pi (X+X')}{2L})|^{1/2}}.
\end{equation}

We can easily generalize this result to obtain the $n$-eigenvalue density matrix:

\begin{equation}
   \begin{split}
        g_n(\{x\},\{x'\})=&\frac{|A_{\psi,\mathcal{V}_{0,-1}}|^{2n}}{\pi^{n/2}} \prod_{i=1}^n\qty|\sin(\pi \frac{X_i}{L})\sin(\pi \frac{X'_i}{L})|^{\frac{1}{4}}\times \\ 
&\frac{\prod_{k<l}\qty|\qty(\frac{2L}{\pi})^2\sin(\pi \frac{(X_k-X_l)}{2L})\sin(\pi \frac{(X_k+X_l)}{2L})\sin(\pi \frac{(X'_k-X'_l)}{2L})	\sin(\pi \frac{(X'_k+X'_l)}{2L})|^{1/2}}{\prod_{i,j}\qty|\frac{2L}{\pi}\sin(\pi \frac{(X_i-X'_j)}{2L}) \sin(\pi \frac{(X_i+X'_j)}{2L})|^{1/2}}.
    \end{split}
\end{equation}

For a double scaled matrix model  $L= \infty$, the one-eigenvalue density matrix is given by:
\begin{equation}
g_1(x,x')=\frac{\sqrt{2}|A_{\psi,\mathcal{V}_{0,-1}}|^2}{\sqrt{\pi}} \frac{|XX'|^{1/4}}{|(X-X')(X+X')|^{1/2}},
\end{equation}
writing it explicitly in terms of the eigenvalues we have:
\begin{equation}
    g_1(E,E')=\frac{\sqrt{2}|A_{\psi,\mathcal{V}_{0,-1}}|^2}{\sqrt{\pi}} \frac{\qty|\qty(\int_{E_L}^E \frac{dE''}{\rho_0(E'')})\qty(\int_{E_L}^{E'} \frac{dE''}{\rho_0(E'')})|^{1/4}}{\qty|\qty(\int_{E}^{E'} \frac{dE''}{\rho_0(E'')})\qty(\int_{E_L}^E \frac{dE''}{\rho_0(E'')}+\int_{E_L}^{E'} \frac{dE''}{\rho_0(E'')})|^{1/2}},
\end{equation}
Approximating the integrals by assuming an almost constant density  $\int_{E_L}^E \frac{dE'}{\rho_0(E')}\approx \frac{E-E_L}{\rho_0(E)}$ we obtain:

\begin{equation}
    g_1(E,E')\approx\sqrt{2}|A_{\psi,\mathcal{V}_{0,-1}}|^2 \frac{\qty|\rho_0(E)\rho_0(E')(E-E_L)(E'-E_L)|^{1/4}}{\qty|E'-E|^{1/2}\qty|E+E'-2E_L|^{1/2}}   .
\end{equation}
In the limit of RMT universality $|E-E'|\ll 1$ we have the simple expression: 

\begin{equation}
    g_1(E,E')\approx|A_{\psi,\mathcal{V}_{0,-1}}|^2 \frac{\qty|\rho_0(E)\rho_0(E')|^{1/4}}{|E-E'|^{1/2}}.
\end{equation}
In the double scaling limit $L=\infty$ the $n$-eigenvalue density matrix is:
\begin{equation}
g_n(\{x\},\{x'\})=|A_{\psi,\mathcal{V}_{0,-1}}|^{2n}\qty(\frac{2}{\pi})^{n/2}\prod_{i=1}^n|X_iX'_i|^{1/4}\times \frac{\prod_{k<l} |(X_k^2-X_l^2)({X'_k}^2-{X'_l}^2)|^{1/2}}{\prod_{i,j}|(X_i^2-{X'}^2_j)|^{1/2}}.
\end{equation}
The factors of $L$ cancel exactly. 

We recognize the Vandermonde determinant $\Delta(X^2)=\prod_{i<j} (X^2_i-X^2_j) $ of the matrix $X^{2(j-1)}_i$:

\begin{equation}
g_n(\{x\},\{x'\})=|A_{\psi,\mathcal{V}_{0,-1}}|^{2n}\qty(\frac{2}{\pi})^{n/2}\prod_{i=1}^n|X_i X'_i|^{1/4}\times \frac{|\Delta(X^2)\Delta({X'}^2)|^{1/2}}{\prod_{i,j}|(X_i-X'_j)(X_i+X'_j)|^{1/2}}.
\end{equation}
To obtain the expression in terms of the eigenvalues it is again enough to substitute $x_i=E_i$ and $X_i=\int_{E_L}^{E_i} \frac{dE}{\pi\rho_0(E)}$. 
Approximating the integrals by assuming an almost constant density  $\int_{E_L}^E \frac{dE'}{\rho_0(E')}\approx \frac{E-E_L}{\rho_0(E)}$ we obtain:
\begin{equation}
\begin{split}
     g_n(\{E\},\{E'\})\approx|A_{\psi,\mathcal{V}_{0,-1}}|^{2n}2^{n/2}&\prod_{i=1}^n \qty|\frac{(E_i-E_L)(E_i'-E_L)}{\rho_0(E_i)\rho_0(E_i')}|^{1/4} \prod_{i,j}\qty|\frac{(E_i-E_L)^2}{\rho_0(E_i)^2}-\frac{(E_j'-E_L)^2}{\rho_0(E_j')^2}|^{-1/2}\\& \prod_{k<l}\qty|\frac{(E_k-E_L)^2}{\rho_0(E_k)^2}-\frac{(E_l-E_L)^2}{\rho_0(E_l)^2}|^{1/2}\qty|\frac{(E_k'-E_L)^2}{\rho_0(E_k')^2}-\frac{(E_l'-E_L)^2}{\rho_0(E_l')^2}|^{1/2}.
\end{split}
\end{equation}
These expressions have been checked against numerical simulations performed via Density Matrix Renormalization Group (DMRG) methods for harmonic and double-well potentials in \cite{brundubaildm}. The hydrodynamic CFT accurately matches the numerical results already for $N=15$ and improves as $N\gg1$.

%% file: 4Discussion.tex
\section{Open questions and future work}
\label{sec:discussion}
We conclude with several questions and possibilities for future work. \\ 

$\bullet$ \textbf{Time-dependent spectral density  $\rho_0(E,t)$:}
We have only considered  fluctuations of the eigenvalues around an equilibrium spectral density  $\rho_0(E)$ which is time independent. In matrix quantum mechanics, the matrix $H(t)$ will evolve in time, thus it is natural to consider a time dependent  density $\rho_0(E,t)$. The hydrodynamic effective theory can  describe such  out-of-equilibrium configurations  and the quantum fluctuations around them. At equilibrium, we have seen that equal time correlations reproduce spectral statistics of matrix integrals such as the one dual to JT gravity.
It would be interesting to understand what is the bulk picture for the time $t$  since it is a priori different from the time $\tau$ coming from analytical continuation of the Euclidean boundary circle $\beta\rightarrow \beta+i\tau$. \\One possibility is that time evolution corresponds to  out-of-equilibrium dynamics in the bulk.
Black hole evaporation is a dynamical out-of-equilibrium process so one might wonder whether we can use matrix quantum mechanics as  a toy model to describe it. We could consider coupling the system to a bath or performing a quench and computing the entanglement entropy as a function of time to see if we have the desired Page curve behaviour. \\ 
Another possibility is to investigate whether it is possible to  dynamically evolve between different theories, e.g. from JT $\rho_0(E,t=0)=\rho_0^{(JT)}(E)$ to a minimal string $\rho_0(E,t^*)=\rho_0^{(2,p)}(E)$ or among minimal strings. This could be interpreted as the matrix model dual of an RG flow between the different minimals models on the worldsheet. Along these lines, it was recently proposed \cite{Suzuki:2021zbe} that  JT gravity is related to the $c=1$ matrix model on a particular time-dependent background.
Since in two dimensional string theory the time $t$  is  simply the time direction in target space \cite{klebanov}, this connection could elucidate the role of $t$.
The effective hydrodynamic approach could be useful for studying time-dependent backgrounds in two-dimensional string theory  as in \cite{Alexandrov:2002fh}. More recently quantum quenches in the $c=1$ matrix model and their string theory interpretation were considered in \cite{Das:2019cgl,Das:2019qaj}. \\

$\bullet$ \textbf{Topological recursion in MQM} 
The duality between JT gravity and a matrix integral was established at all orders in $\frac{1}{N}$ thanks to topological recursion \cite{Saad:2019lba, Eynard:2015aea, mirza}. It would then be good to understand  topological recursion from the point of view of MQM. In particular, the  $\frac{1}{N}\sim \hbar$  corrections in MQM are given by higher orders in the WKB expansion of the eigenvalue wavefunction $\psi(E)$. It has been shown, for certain classes of spectral curves, that the WKB expansion of an associated quantum mechanical system satisfies topological recursion \cite{Bouchard:2016obz}.\footnote{The class of spectral curves for which this has been shown does not include JT gravity's spectral curve but it does include the Airy case $\rho(E)=\sqrt{E}$.} This connection between WKB and topological recursion might shed  light on MQM and its \textit{one-time-point} reduction to the matrix integral dual to JT gravity. \\

$\bullet$ \textbf{Universe field theory:}
We can think of  Matrix Quantum Mechanics  as a  quantization of the matrix integral dual to JT gravity. We define an operator $\hat{Z}(\beta)$ given by: 
\begin{equation}
  \hat{Z}(\beta)=\int dE e^{-\beta E} \hat{\rho}(E)
\end{equation} 
which creates a spacetime with a boundary of length $\beta$. 
$\hat{Z}(\beta)$ gives a realization of the operators acting on the Hilbert space of baby universes  discussed in \cite{Marolf:2020xie,Post:2022dfi,Altland:2022xqx}.\\ In particular, understanding the connection between the universe field theory defined by  MQM and the one proposed in \cite{Post:2022dfi,Altland:2022xqx} is an interesting prospect which  deserves further investigation. From this point of view, the time $t$ in MQM corresponds to the time evolution $e^{i\mathcal{H}_{BU}t}$ generated by the third quantised Hamiltonian $\mathcal{H}_{BU}$ on the baby universe Hilbert space. This Hamiltonian should then be identified with the Hamiltonian of the MQM in eq. \ref{eq:hamiltonianmqm}.\\
Similarly to $\hat{Z}(\beta)$, the eigenvalue wavefunction $\psi_N(E_1\dots E_N)$ and the reduced density matrix $g_n(E,E')$ describe the Wheeler–DeWitt wavefunction of universes with specified boundaries.  It would be interesting to understand better the implications of the reduced density matrix in this context.
 In  two-dimensional string theory the operators $\hat{Z}(\beta)$ are known as loop operators and their third quantised interpretation in the $c=1$ matrix model  has been discussed in \cite{Betzios:2020nry}.   \\

$\bullet$ \textbf{Finite temperature and non-singlet sector:} 
We can consider matrix quantum mechanics at finite temperature by compactifying the time direction $t$ with period $2\pi R$ (see secs. 8, 9 and 10 of \cite{klebanov}). It is well known that a   Berezinskii–Kosterlitz–Thouless (BKT) phase transition takes place: for $R<R_{BKT}$ vortices condense and the non-singlet degrees of freedom dominate the free energy \cite{Gross:1990md,Gross:1990ub,Yang:1990ns, Boulatov:1991xz}. The physics of the non-singlet sector is very rich, involving 2D black holes and long strings \cite{Karczmarek:2004bw,Kazakov:2000pm,Maldacena:2005hi}. Thus it would be interesting to understand the transition by incorporating vortices into the hydrodynamic effective theory. 
Moreover, at high temperatures $R\rightarrow 0$, fluctuations along the thermal circle are suppressed and we recover a 0-dimensional matrix integral with potential $V(H)$. This suggests that we could think of the matrix integral dual to JT gravity as a high temperature limit of Matrix Quantum Mechanics. From this point of view, JT gravity would be dual to  a single quantum  mechanical system in the high temperature limit. The apparent ensemble averaging could be due to the system being in a disordered phase at high temperature.

\section*{Acknowledgements}
We would like to thank Bruno Balthazar, Shaun Hampton, Vladimir Kazakov, Pierfrancesco Urbani and Takato Yoshimura for useful discussions and comments on the manuscript.  GDU's research is supported by ERC Starting Grant 853507. GDU would like to thank École Normale Supérieure, where this project started, for previous support in the form of a LABEX ENS-ICFP scholarship.